\documentclass{aa}

\usepackage{graphicx}
\usepackage[varg]{txfonts}
\usepackage{natbib}
\usepackage{longtable}
\usepackage{array}
\usepackage{multirow}
\usepackage{txfonts}

\usepackage{dcolumn}
\newcolumntype{k}[1]{D{.}{.}{#1}}
\newcolumntype{d}{D{.}{.}{-1}}
\newcolumntype{,}{D{,}{,}{-1}}

\bibpunct{(}{)}{;}{a}{}{,}
\begin{document}

\title{Millimetre-wave laboratory study of glycinamide and search for it with ALMA toward Sagittarius B2(N)
\thanks{Tables 5, 6, and 7 are only available in electronic form
at the CDS via anonymous ftp to cdsarc.u-strasbg.fr (130.79.128.5) or via http://cdsweb.u-strasbg.fr/cgi-bin/qcat?J/A+A/
}}

   \author{Z. Kisiel\inst{1}
      \and L. Kolesnikov\'a\inst{2,3}
      \and A. Belloche\inst{4}
      \and J.-C. Guillemin\inst{5}
      \and L. Pszcz\'o{\l }kowski\inst{1}
      \and E. R. Alonso\inst{3}
      \and R. T. Garrod\inst{6}
      \and E. Bia{\l }kowska-Jaworska\inst{1}
      \and I. Le\'{o}n\inst{3}
      \and H.S.P. M\"uller\inst{7}
      \and K.M. Menten\inst{4}
      \and J. L. Alonso\inst{3}
          }

\institute{Institute of Physics, Polish Academy of Sciences. Al. Lotnik\'ow 32/46, 02-668 Warszawa, Poland
\and       Department of Analytical Chemistry, University of Chemistry and Technology, Technick\'a 5, 166 28, Prague 6, Czech Republic
\and       Grupo de Espectroscop\'ia Molecular (GEM), Edificio Quifima, \'Area de Qu\'imica-F\'isica, Laboratorios de Espectroscop\'ia y
Bioespectroscop\'ia, Parque Cient\'ifico UVa, Unidad Asociada CSIC, Universidad de Valladolid, 47011 Valladolid, Spain
\and       Max-Plack-Institut f\"ur Radioastronomie, Auf dem H\"ugel 69, 5321 Bonn, Germany
\and       Univ Rennes, Ecole Nationale Sup\'erieure de Chimie de Rennes,
CNRS, ISCR-UMR6226, 35000 Rennes, France
\and       Departments of Chemistry and Astronomy, University of Virginia, Charlottesville, VA 22904, USA 
\and       Astrophysik\ I. Physikalisches Institut, Universit{\"a}t zu K{\"o}ln, Z{\"u}lpicher Str. 77, 50937 Cologne, Germany \\
\email{kisiel@ifpan.edu.pl (ZK) \& belloche@mpifr-bonn.mpg.de (AB)}
           }

\date{Received 1 October 2021; accepted 17 October 2021}

\titlerunning{The rotational spectrum and ISM search for glycinamide}
\authorrunning{Kisiel et. al.}


\abstract
{Glycinamide (NH$_2$CH$_{2}$C(O)NH$_{2}$) is considered to be one of the
possible precursors of the simplest amino acid glycine. Its only
rotational spectrum reported so far has been in the cm-wave region on a
laser-ablation generated supersonic expansion sample.}
{The aim of this work is to extend the laboratory spectrum of glycinamide
into the millimetre wave region to support its searches in the interstellar medium and
to perform the first check for its presence in the high-mass star forming region Sagittarius B2(N).}
{Glycinamide was synthesised chemically and was studied with
broadband rotational spectroscopy in the 90-329 GHz region with the
sample in slow flow at 50$^\circ$C. Tunneling across a low energy barrier
between two symmetry equivalent configurations of the molecule resulted in splitting of each vibrational
state and many perturbations in associated rotational energy levels, requiring careful
coupled state fits for each vibrational doublet.
We searched for emission of glycinamide in the imaging spectral line survey 
ReMoCA performed with the Atacama Large Millimetre/submillimetre Array toward 
Sgr B2(N). The astronomical spectra were analysed under the assumption
of local thermodynamic equilibrium.}
{We report the first analysis of the mm-wave rotational spectrum of glycinamide, resulting
in fitting to experimental measurement accuracy of over 1200 assigned and measured transition frequencies
for the ground state tunneling doublet, of many lines for tunneling doublets for two singly excited vibrational states,
and determination of precise vibrational separation in each doublet.
We did not detect emission from glycinamide in the hot molecular core
Sgr~B2(N1S). We derived a column density upper limit of 
$1.5 \times 10^{16}$~cm$^{-2}$, which implies that glycinamide is at least
seven times less abundant than aminoacetonitrile and 1.8 times less abundant 
than urea in this source.
A comparison with results of astrochemical kinetics models for species 
related to glycinamide suggests that its abundance may be at least one order of
magnitude below the upper limit obtained toward Sgr~B2(N1S). This means that
glycinamide emission in this source likely lies well below the spectral 
confusion limit in the frequency range covered by the ReMoCA survey.
   }
{Thanks to the spectroscopic data provided by this study, the search 
for glycinamide in the interstellar medium can continue on a firm 
basis. Targetting sources with a lower level of spectral confusion, 
such as the Galactic Center shocked region G+0.693-0.027, may be a 
promising avenue.}


   \maketitle
%

\section{Introduction}
\label{sect_intro}

The quest for the simplest amino acid glycine in the interstellar medium
(ISM) became a never-ending story for researches in the fields of
astrochemistry and astrophysics. Since its first interstellar hunt
\citep{Brown1979} more than forty years have passed during which glycine has
been extensively searched for toward various interstellar sources in
both the centimetre and the millimetre wave regions of the electromagnetic
spectrum
\citep{Hollis1980,Snyder1983,Berulis1985,Guelin1989,Combes1996,
Ceccarelli2000,Hollis2003,Kuan2003,Kuan2004,Belloche2008}. However, its presence
in the ISM has never been confirmed
\citep{Snyder2005,Jones2007,Cunningham2007}, neither in the era of ALMA.
This is in spite of the fact that glycine has been discovered in
meteorites \citep{Pizzarello1991,Ehrenfreund2001,Glavin2006}, dust
samples from comet Wild~2 \citep{Elsila2009}, and in the coma of comet
67P/Churyumov-Gerasimenko  \citep{Altwegge2016}. In addition, several
laboratory experiments demonstrated the synthesis of glycine, and other
amino acids, when interstellar ice analogs were subjected to UV
radiation \citep{Bernstein2002,Munoz2002,Lee2009,Zheng2010,Kim2011} or
bombarded by energetic electrons \citep{Holtom2005}.
Numerous studies were undertaken to throw light on this controversy.
They focused on plausible interstellar pathways to glycine, its
detectability and survival in the hostile ISM \citep[see,
e.g.,][]{Ehrenfreund2001a,Blagojevic2003,Largo2010,Pilling2011,
Lattelais2011,Rimola2012,Garrod2013,Serra2014,Nhlabatsi2016,Aponte2017,Suzuki2018,
Xavier2019}. Special emphasis has been also placed on the formation
routes of possible glycine precursors \citep[see, e.g.,][and references
therein]{Basiuk2001,Largo2004,Koch2008,Knowles2010,Barrientos2012,Redondo2015}
which are hot candidates for the observations in the ISM as well.
Some of them, such as methylamine \citep{Kaifu1974}, aminoacetonitrile
\citep{Belloche2008}, and hydroxylamine \citep{Rivilla2020} have been
already detected. Rotational spectroscopic studies of potential glycine
precursors hydantoin \citep{Alonso2017,Ozeki2017} and hydantoic acid
\citep{Kolesnikova2019} were reported recently. \cite{SanzNovo2019}
further computed the spectroscopic properties of glycine isomers of
which methyl carbamate
\citep{Marstokk1999,Bakri2002,Ilyushin2006,Groner2007} and glycolamide
\citep{Maris2004,SanzNovo2020} were studied by microwave and millimetre
wave spectroscopies that enabled their searches in the ISM \citep{SanzNovo2020,Sahu2020}.

In this work we focus on glycine precursor glycinamide
(NH$_2$CH$_{2}$C(O)NH$_{2}$) which is predicted to be a feasible
intermediate on the hydrolytic way from aminoacetonitrile to glycine
\citep{Zhu2004,Ugliengo2011}. Keeping in mind the presence of
aminoacetonitrile in Sgr B2(N) \citep{Belloche2008}, glycinamide could
be considered as a good candidate for observations in the same source.
Millimetre wave surveys of high-mass star-forming regions are known to
present a forest of lines with a high level of line blending \citep[see
e.g.][]{Tercero2010,Belloche2013}. For this reason, the identification
of a new molecule, such as glycinamide, in these sources has to be
guaranteed by the detection of numerous features consistent with
confident predictions of its spectrum over a broad frequency region. High-quality laboratory
data and analysis are thus the first and mandatory step before any
interstellar search can be conducted.

\begin{figure}[t]
\centering
\includegraphics[width=7.5cm]{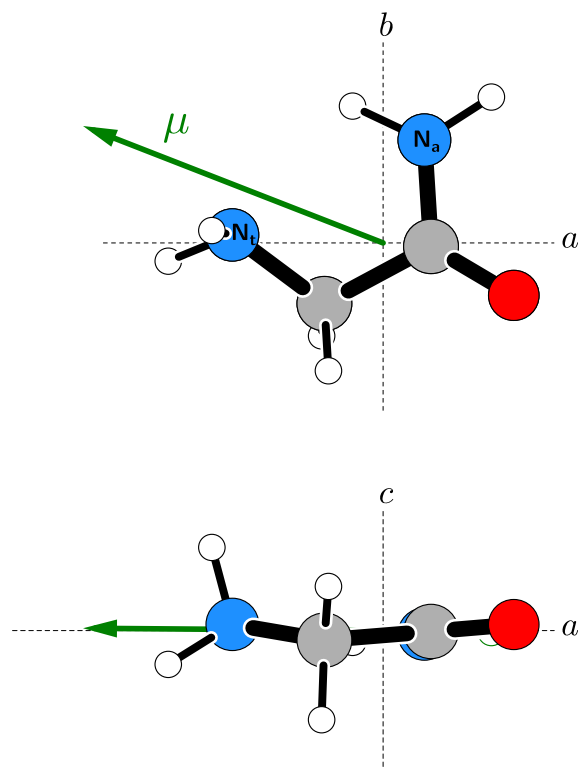}
\label{fig:struct}

\caption{Geometry and electric dipole moment of the most stable conformer
of glycinamide, and their orientation in the principal rotational axis frame.  The
dipole moment vector is confined to the $ab$ inertial plane, and is drawn from the notional
negative to the notional positive charge.  The molecule has a
planar amido N$_a$H$_2$ group, and a pyramidal amino N$_{\rm t}$H$_2$ group. The
geometry is non-planar with a small $\sim$ 100 cm$^{-1}$ barrier at the
$C_s$-symmetry configuration, where all heavy atoms are in the same plane, which also
bisects the CH$_2$ and N$_{\rm t}$H$_2$ groups. Tunneling across the barrier
between the two equivalent non-planar configurations leads to
appreciable splitting into pairs of all vibrational states, including the ground state. }

\end{figure}

It was only recently that the rotational spectrum of glycinamide was
first studied. \cite{Alonso2018} investigated its conformational
landscape in supersonic expansion by Fourier transform microwave
spectroscopy between 6~GHz and 16~GHz. The analysis of the spectrum
revealed the existence of a single conformer whose configuration in the
principal axis frame is shown in Fig.~\ref{fig:struct}. In addition to
this, an unexpected non-rigid behavior of this conformer has been
implied by abnormal values of quartic centrifugal
distortion constants. Such a behavior has been attributed to a
large-amplitude motion that combines C–N$_{\text{t}}$ bond torsion, C-C torsion, and
N$_{\text{t}}$H$_{2}$ inversion. Since this motion is governed by tunneling through
the central barrier in a
double minimum potential function, rotational transitions in the ground
vibrational state are expected to be split into two components
associated with two torsion/inversion sublevels,  usually labeled $0^{+}$ and $0^{-}$.
However, due to vibrational cooling accompanying supersonic expansion, only the transitions in the
lowest-lying $0^{+}$ substate were observed by \cite{Alonso2018}. For
interpreting the dense millimetre wave surveys from interstellar
sources, the rotational transitions not only in this $0^{+}$ substate
but also in the yet experimentally unobserved $0^{-}$ substate might be of
critical importance. Furthermore, severe perturbations resulting from
the $0^{+} \leftrightarrow 0^{-}$ coupling of the rotational manifolds
might be manifested for higher $J$ and $K_{a}$ transitions in the
millimetre wave region. This situation is actually the case for cyanamide
\citep{Read1986,Krasnicki2011,Kisiel2013,Coutens2019} in which the large-amplitude
motion involves the inversion of the NH$_{2}$ group. The rotational
transitions of cyanamide in the $0^{+}$ and $0^{-}$ substates could be
detected towards solar-type protostars \citep{Coutens2018} only after a
successful laboratory analysis accounting for the $0^{+} \leftrightarrow
0^{-}$ interactions \citep{Read1986,Krasnicki2011,Kisiel2013}. From this
it turns out that common difficulties arising from extrapolations to
higher frequencies are not the only problems faced by attempts to
detect glycinamide in space.

In the course of the present work, the laboratory rotational spectrum of glycinamide
has been measured between 90 GHz and 329 GHz. A comprehensive analysis
of this millimetre wave spectrum made it possible to unambiguously
assign the rotational transitions in the four lowest torsional/inversion substates
$0^{+}$, $0^{-}$, $1^{+}$ and $1^{-}$. In addition, rotational
transitions in $0^{+}$ and $0^{-}$ substates for $v_{26}=1$ state were
assigned and measured. Results of this work deliver first experimental
information on the double minimum potential of glycinamide and provide a
firm basis for its searches in space.

Section 2 describes the laboratory experiments performed to measure the
rotational spectrum of glycinamide. The analysis of the laboratory spectra is 
presented in Section 3 and the search for glycinamide toward the hot molecular 
core Sgr~B2(N1) with the Atacama Large Millimetre/submillimetre 
Array (ALMA) is reported in Section 4. Section 5 
discusses the spectroscopic and astronomical results, and Section 6 
summarises our conclusions.

%

\begin{figure*}[!ht]
\centerline{\resizebox{1.0\hsize}{!}{\includegraphics[angle=-90]{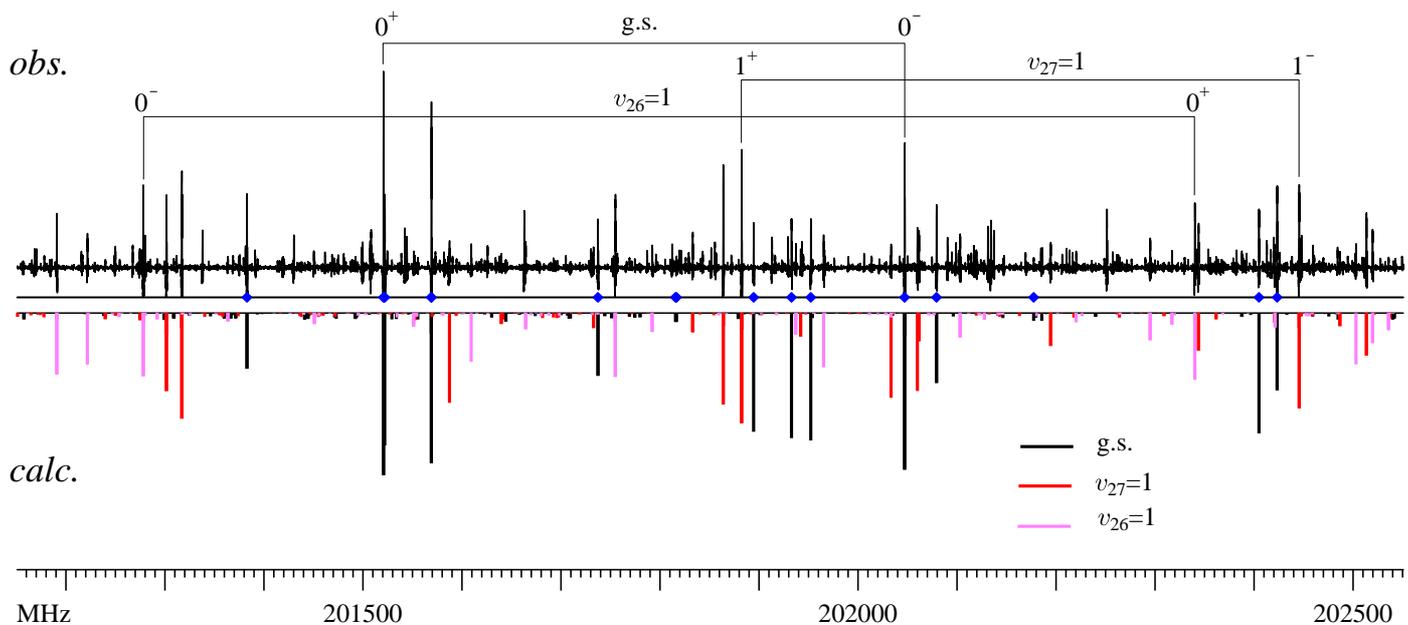}}}

\caption{Sample 1.4 GHz of the rotational spectrum of
glycinamide at 50$^\circ$C illustrating the absence of characteristic spectral patterns.
Overbars in the experimental spectrum (top panel) mark $J = 34 \leftarrow 33$ transitions 
between rotational levels with
$K_a$=0 and 1, with each marked line corresponding to an overlap of a pair of
stronger $\mu_a$-dipole, and a pair of weaker $\mu_b$-dipole transitions. The frequency
span of the bar indicates the significant magnitude of the splitting between
two substates for each of the three tunneling doublets. Blue
markers indicate ground state transitions in the data set used for determining
spectroscopic constants, while the bottom panel displays collected predictions made on 
the basis of the final fits.}\label{fig:mmw}

\end{figure*}

\section{Experiments}
{\label{s:experiments}}

Pure glycinamide has the form of colourless crystals with a melting point of
63$^\circ$C.  A vapour pressure of 5-10 mTorr that was sufficient for measurements was
generated when the sample was heated to 50$^\circ$C, and signal stability was ensured
by using a slow flow through the 3-metre free space absorption cell of the spectrometer.
In the previous rotational spectroscopy study \citep{Alonso2018} glycinamide was produced by laser ablation
of a specially prepared hydrochloride glycinamide rod, while presently we used pure glycinamide.
This was prepared from glycinamide hydrochloride, as purchased from Aldrich and used without
further purification. Glycinamide hydrochloride (5.5 g, 50 mmol) and dry
dichloromethane (100 mL) were introduced under nitrogen into a 250 ml
three-necked flask. Dry ammonia was then bubbled through the stirred
suspension for 30 min. The salt was filtered off and the solvent was
removed under vacuum on a rotary evaporator and then on a vacuum pump
(0.1 mbar, 2 h). Glycinamide (2.5 g, 34 mmol, 68\%\ yield) was thus
obtained.  It showed long term stability when stored in the freezer (-20$^{\circ}$C) and
there was also no discernible decomposition when heated for measurements.

Spectroscopic measurements were made in the region 90$-$329~GHz with the broadband backward wave oscillator (BWO) based
spectrometer in Warsaw, which was described in \cite{medvedev2004}.  The hardware configuration was
later updated and augmented
below 140 GHz with a harmonic generation source \citep{Kisiel2010}.  
A frequency measurement accuracy of 50 kHz was assumed.

\section{Rotational spectra and analysis}
{\label{s:analysis}}

The mm-wave rotational spectrum of glycinamide posed a significant assignment
challenge, which can be partly appreciated from the example shown in
Fig.~\ref{fig:mmw}. This is a rather asymmetric molecule (asymmetry parameter
$\kappa=-0.68$), expected to give rise to strong pure rotation $a$-type transitions accompanied by
weaker $b$-type transitions (calculated dipole moment components are
$\mu_a$=3.8 D and $\mu_b$=1.5 D, \citealt{Alonso2018}). There is, however, a
significant lack of characteristic line patterns that normally aid assignment.
The most useful starting point was provided by the fact that the strongest
lines in the spectrum would be quadruply degenerate by resulting from overlaps of
pairs of $a$-type and $b$-type  $R$-branch transitions for the lowest values of
$K_a$.  Unfortunately such lines quickly proved to be poorly treatable with a
single state Watson's asymmetric rotor Hamiltonian \citep{Watson1977}.  On the other
hand, several sequences of such strong lines were identified with the use of the AABS graphical assignment
package \citep{aabs,Kisiel2012}.  Initial rotational constants, reported by \cite{Kisiel2010a},
were confirmed by measurement of the lowest-$J$ ground state transitions in supersonic
expansion \citep{Alonso2018}.

Complete assignment and analysis of the mm-wave spectrum became possible
once it was realised that the observed sets of transitions belonged to
coupled pairs of vibrational sublevels. As already mentioned, such pairs of levels typically
arise through tunneling between two symmetry equivalent structures of the
molecule that are separated by a relatively low energy barrier. Two
clear boundary conditions are possible, where tunneling is equivalent to
a torsional motion (as in phenol, \citealt{Kol1}), or to inversion (as in
cyanamide, \citealt{Kisiel2013}).  For glycinamide computations indicated
that tunneling between the most stable conformation depicted in
Fig.~\ref{fig:struct} and its mirror image form obtained by reflection
across the $ab$ inertial plane is subject to a barrier of only $\sim$
100~cm$^{-1}$  \citep{Alonso2018} and that such tunneling has to involve
a concerted combination of torsion and inversion.  The molecule belongs
to the lowest symmetry point group, $C_1$, and the general labeling used
for vibrational levels split by double minimum inversion has been
adopted (see Fig.~25.2 of \citealt{PapAliev}).  Accordingly, the
vibrational ground state, normally labeled $v=0$, splits into a pair of
sublevels designated 0$^+$ (lower) and 0$^-$ (upper).  Similarly, the
usual first excited state, $v=1$, of the inversion motion, becomes a
doublet labeled $1^+$ (lower) and $1^-$ (upper).  This splitting structure is
expected to be relatively unaffected by excitation of non-inverting vibrational
motions.  Accordingly, the first excited state of such a motion,
for example labeled $v_a=1$, will be split by an amount comparable to that in the ground state, with
labeling $v_a=1, 0^+$ (lower) and $v_a=1, 0^-$ (upper).  Quantum chemistry
vibrational calculations
at both harmonic \citep{Li2003} and anharmonic \citep{Alonso2018} levels
indicated, in agreement, that glycinamide has two low frequency vibrational modes $\nu_{27}$ and
$\nu_{26}$ with vibrational frequencies of near 100 and 200~cm$^{-1}$, respectively.
Nevertheless, unambiguous attribution as to which of these modes is responsible for the tunneling
only became possible on completion of the rotational analysis, as discussed below.

The most effective treatment of pairs of vibrational substates resulting from
tunneling is to use a (2$\times$2) block diagonal Hamiltonian of the form:

\begin{eqnarray}
 H =       \left(
              \begin{array}{cc}
            H_{\rm rot}^{(0^+)}                        &    H_{\rm c}^{(0^+,0^-)}   \\
            \\
            H_{\rm c}^{(0^+,0^-)}                       &    H_{\rm rot}^{(0^-)}+\Delta E
              \end{array}
                                                      \right).
\label{eqham}
\end{eqnarray}
%
%


\begin{table*}[!h]
\caption{Spectroscopic constants determined for the three assigned tunneling doublets in glycinamide.}
\vspace{-0.5cm}
\label{tab:const}
\renewcommand{\thefootnote}{\alph{footnote}}
\begin{center}
\begin{footnotesize}
\begin{tabular}{lk{9}k{9}k{9}k{9}k{9}k{9}}
\hline\hline\vspace{-0.2cm}\\
                                 & \multicolumn{2}{c}{ground state}
                                 & \multicolumn{2}{c}{$v_{27}=1$}
                                 & \multicolumn{2}{c}{$v_{26}=1$} \\
Parameter                        & \multicolumn{1}{c}{ $0^+$}
                                 & \multicolumn{1}{c}{ $0^-$}
                                 & \multicolumn{1}{c}{ $ 1^+$}
                                 & \multicolumn{1}{c}{ $ 1^-$}
                                 & \multicolumn{1}{c}{ $ 0^+$}
                                 & \multicolumn{1}{c}{ $ 0^-$} \\
\vspace{-0.2cm}\\
\hline
\\
$A^{a}$  /MHz                   &   9631.65562(72)^b  & 9609.18503(85)      &  9592.9536(13)    &   9560.8227(15)    &   9561.1846(27)     &   9616.0695(33)    \\
$B    $  /MHz                   &   3986.80416(25)    & 3992.03664(20)      &  3984.61628(30)   &   3992.78766(28)   &   3991.12576(35)    &   3978.65176(43)   \\
$C    $  /MHz                   &   2925.57754(23)    & 2931.88720(21)      &  2930.85963(37)   &   2939.30567(39)   &   2937.87100(20)    &   2921.79555(51)   \\
                                &                     &                     &                   &                    &                     &                    \\
$\Delta_{J} $  /kHz  ~~~~       &      0.82062(13)    &    0.79899(14)      &     0.75262(19)   &      0.81389(20)   &      0.80404(11)    &      0.74937(27)   \\
$\Delta_{JK}$  /kHz             &      3.13182(86)    &    3.42634(91)      &     4.338(12)     &      3.307(12)     &      4.1747(17)     &      6.2857(24)    \\
$\Delta_{K} $  /kHz             &      4.5931(23)     &    4.0563(26)       &     2.635(13)     &      4.562(16)     &      4.890(16)      &      0.222(14)     \\
$\delta_{J} $  /kHz             &      0.182640(96)   &    0.186058(37)     &     0.185750(58)  &      0.177310(54)  &      0.163415(58)   &      0.190534(89)  \\
$\delta_{K} $  /kHz             &      2.65419(92)    &    1.99562(87)      &     3.0722(53)    &      1.2000(50)    &      2.8152(11)     &      0.2459(43)    \\
                                &                     &                     &                   &                    &                     &                    \\
$\Phi_{J} $    /Hz              &      0.000185(37)   &    0.000505(39)     &     0.000407(43)  &      0.000244(46)  &    [ 0.]            &      0.001143(64)  \\
$\Phi_{JK}$    /Hz              &     -0.05630(39)    &    0.01800(100)     &    -0.1036(42)    &      0.0869(42)    &      0.0225(15)     &     -0.0342(35)    \\
$\Phi_{KJ}$    /Hz              &     -0.0226(27)     &   -0.0409(33)       &     0.0534(81)    &     -0.2674(82)    &     -0.4170(57)     &     -0.830(12)     \\
$\Phi_{K}$     /Hz              &      0.0591(28)     &  [ 0.]              &     0.0765(81)    &      0.082(14)     &      0.339(26)      &      0.602(22)     \\
$\phi_{J}$     /Hz              &     -0.000308(27)   &  [ 0.]              &   [ 0.]           &    [ 0.]           &    [ 0.]            &    [ 0.]           \\
$\phi_{JK}$    /Hz              &    [ 0.]            &  [ 0.]              &   [ 0.]           &    [ 0.]           &    [ 0.]            &      0.0196(16)    \\
$\phi_{K}$     /Hz              &    [ 0.]            &   -0.1395(70)       &    -0.055(11)     &      0.076(10)     &      0.181(11)      &     -0.318(27)     \\
                                &                     &                     &                   &                    &                     &                    \\
$\Delta E$     /MHz             &                     & 287355.66(10)       &                   &  982174.88(42)     &                     & 304943.46(32)      \\
$\Delta E$     /cm$^{-1}$       &                     &      9.585186(3)    &                   &      32.761828(14) &                     &     10.171818(11)  \\
                                &                     &                     &                   &                    &                     &                    \\
$F_{bc}$       /MHz             &                     &      4.2369(31)     &                   &      11.573(17)    &                     &      1.1549(28)    \\
$F_{bc}^J$     /kHz             &                     &      0.3060(12)     &                   &       0.3553(75)   &                     &     -0.07722(63)   \\
$F_{bc}^K$     /kHz             &                     &      0.457(31)      &                   &     [ 0.]          &                     &      2.692(21)     \\
$F_{bc}^{JJ}$  /Hz              &                     &      0.01487(28)    &                   &     [ 0.]          &                     &    [ 0.]           \\
$F_{bc}^{JK}$  /Hz              &                     &      0.1352(90)     &                   &     [ 0.]          &                     &    [ 0.]           \\
$F_{bc}^{KK}$  /Hz              &                     &     -0.339(69)      &                   &     [ 0.]          &                     &    [ 0.]           \\
                                &                     &                     &                   &                    &                     &                    \\
$F_{ca}$       /MHz             &                     &     52.87384(66)    &                   &      35.424(84)    &                     &      3.4185(65)    \\
$F_{ca}^J$     /kHz             &                     &     -0.40174(78)    &                   &      -1.309(25)    &                     &     -0.2358(71)    \\
$F_{ca}^K$     /kHz             &                     &      2.8898(46)     &                   &       4.460(44)    &                     &      2.183(34)     \\
$F_{ca}^{JJ}$  /Hz              &                     &     -0.00668(17)    &                   &      -0.0416(22)   &                     &     -0.0114(19)    \\
$F_{ca}^{JK}K$ /Hz              &                     &     -0.0450(22)     &                   &       0.2102(99)   &                     &      0.1119(75)    \\
$F_{ca}^{KK}K$ /Hz              &                     &      0.2775(79)     &                   &     [ 0.]          &                     &     -0.222(31)     \\
                                &                     &                     &                   &                    &                     &                    \\
$N_{\text{lines}}^{c}    $      &                     &   1237              &                   &      873           &                     &     714            \\
$\sigma_{\text{fit}}^{d} $ /kHz &                     &     43.76           &                   &       44.32        &                     &      50.31         \\
$\sigma_{\text{rms}}^{e} $      &                     &      0.8865         &                   &        0.8863      &                     &       1.0056       \\
\\
\hline
\end{tabular}
\end{footnotesize}
\end{center}
{\bfseries Notes.}\\
$^{(a)}$$A$, $B$, and $C$ are rotational constants, $\Delta_{J},..., \phi_{K}$ are centrifugal distortion constants in Watson's $A$-reduced
asymmetric rotor Hamiltonian, $\Delta E$ is the fitted energy difference between the two levels of the tunneling doublet,
$F_{bc}$ and $F_{ca}$ are leading substate coupling terms in the RAS Hamiltonian, and the related superscripted versions are
parameters in their empirical centrifugal distortion expansions.
$^{(b)}$Errors in parentheses are standard errors in units of the last digit.
$^{(c)}$Number of fitted lines.
$^{(d)}$Standard deviation of the fit.
$^{(e)}$Unitless (weighted) deviation of the fit.
\vspace{0.5cm}
\end{table*}

\noindent The two diagonal blocks, $H_{\rm rot}^{(0^+)}$ and $H_{\rm rot}^{(0^-)}$ are set up with the standard
asymmetric rotor Hamiltonian \citep{Watson1977} for each of the two substates.
The two off-diagonal blocks, $H_{\rm c}^{(0^+,0^-)}$, connecting the two substates are set up with the
Reduced Axis System (RAS) Hamiltonian \citep{Pickett1972}.  In the case where the tunneling motion is around an axis
in the $ab$ inertial plane, but at some angle to the principal axes, the RAS blocks are:

\begin{eqnarray}
 H_{\rm c}^{(0^+,0^-)}  & = &  (F_{bc}+ F_{bc}^J P^2 + F_{bc}^{K} P_z^2 +\dots) ( P_b P_c + P_c P_b) + \nonumber\\
                        &   &  (F_{ca}+ F_{ca}^J P^2 + F_{ca}^{K} P_z^2 +\dots) ( P_c P_a + P_a P_c),
\label{eqcor}
\end{eqnarray}

\noindent where $F_{bc}$ and  $F_{ca}$ are the main adjustable parameters describing the interaction, each of which is
further expanded empirically using centrifugal distortion type terms $F_{bc}^J$, $F_{bc}^{K}$, etc.
Finally, $\Delta E$ is the vibrational energy difference between the two substates, $E(0^-)-E(0^+)$.
This approach was used, for example, for the singly deuterated species, HDNCN, of
cyanamide \citep{Kisiel2013}.
All fits and predictions were carried out with Pickett's SPFIT/SPCAT package \citep{Pickett1991}.

Assignment proceeded through analysis of $R$-branch transition sequences
for successively higher values of $K_a$.  Once these were understood for
the ground state tunneling doublet, it was possible to also assign two
vibrationally excited doublets. The success of the coupled state fit for each such pair of substates was critically
dependent on the value of  $\Delta E$.  In each case, the range of likely values had to be scanned with some care, prior to
using it as an adjustable parameter of fit.

The final understanding of the spectrum
is illustrated in the lower part of Fig.~\ref{fig:mmw}.  It can be seen
from the pairs of lowest $K_a$ transitions marked in the top of
Fig.~\ref{fig:mmw} that the frequency differences between the same rotational transitions in
the substates
are considerable, which reflects the relatively low barrier to the tunneling.
Extensive relative intensity measurements allowed the energies above the
ground state for the two excited vibrational states (energy differences between the lower substates) to be determined as
99(13) and 201(13) cm$^{-1}$, which has been the initial basis for the $v_{27}=1$ and
$v_{26}=1$ assignment marked in Fig.~\ref{fig:mmw}.

\begin{figure}[t]
\centering
\includegraphics[width=7.0cm]{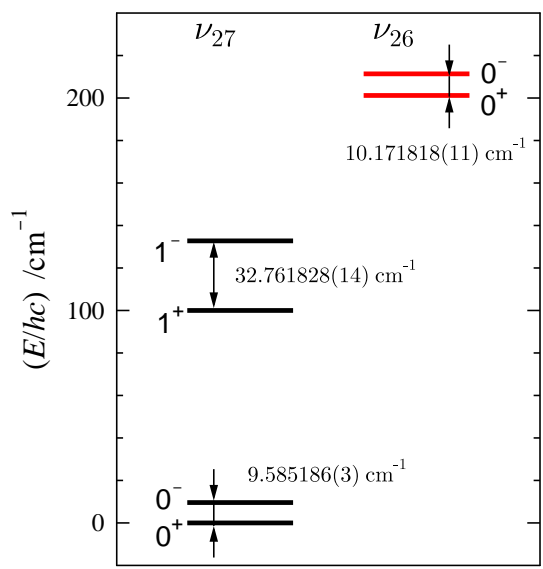}

\caption{Vibrational energies of the three tunneling doublets studied in this work.
Each splitting is determined precisely from the coupled fits of measured rotational frequencies, while
relative positions of the two vibrationally excited doublets are determined less precisely
from relative intensity measurements.}\label{fig:elevels}

\end{figure}

Spectroscopic constants resulting from the fits are summarised in
Table~\ref{tab:const} and a breakdown of statistics for individual
substates is given in Table~\ref{tab:subsets}. Data for the $0^+$ substate
also include hyperfine removed frequencies from supersonic expansion
measurements reported in \cite{Alonso2018}.  For the two ground state
substates some $Q$-branch transitions were also assigned and measured.  In
addition, nominal interstate transitions were observed.  These result from
strong mixing between perturbing rotational levels in two substates and are usually
only identified in the final stages of the analysis.

For all of the identified substates, as well as for combined results for
their pairs, it was possible to reproduce measured frequencies to within
their assumed experimental accuracy of 50~kHz. The energy differences
between the tunneling substates have been determined very precisely, to
sub-MHz precision.  When combined with results of relative intensity
measurements these allow determination of the positions of lowest
vibrational energy levels in glycinamide, as summarised in
Fig.~\ref{fig:elevels}.  We now have further evidence concerning the proposed
vibrational assignment.  Tunneling splitting is expected to
increase significantly with vibrational excitation in a double minimum
potential and can be
successfully modeled with simple potentials, as described and carried
out for cyanamide \citep{Kisiel2013}.  The two splitting values in the
$\nu_{27}$ column in Fig.~\ref{fig:elevels} are consistent with a relatively low barrier
so that this mode is responsible for the tunneling.  On the other hand, the
splitting in the $\nu_{26}$ tunneling pair is very close to that in the
ground state, so that this appears to be a standard normal vibrational mode.

The coverage of values of rotational quantum numbers
by the measured transitions is rather comprehensive and
can be assessed from the data distribution plots for
the three studied doublets given in Figs.~\ref{fig:distr0} -- \ref{fig:distr2}.
The values of quartic centrifugal distortion constants in the substates in each
tunneling pair are generally quite close to each other, which shows that the coupling behaviour
has been largely accounted for by the parameters in Eq.~\ref{eqcor}.
Two significant exceptions are $\Delta_K$ and $\delta_K$ for the upper substate of $v_{26}=1$,
suggesting that this substate may be interacting with some state outside the model, possibly $2^{+}$
of $\nu_{27}$.
The dominance in magnitude of the $F_{ca}$ over the $F_{bc}$ parameter is
similar to that for HDNCN \citep{Kisiel2013} and can accordingly be
taken to be an indicator that the effective axis around which tunneling
takes place is at a relatively small angle to the $b$-inertial axis.

\begin{figure}[t]
\centering
\includegraphics[width=8.0cm]{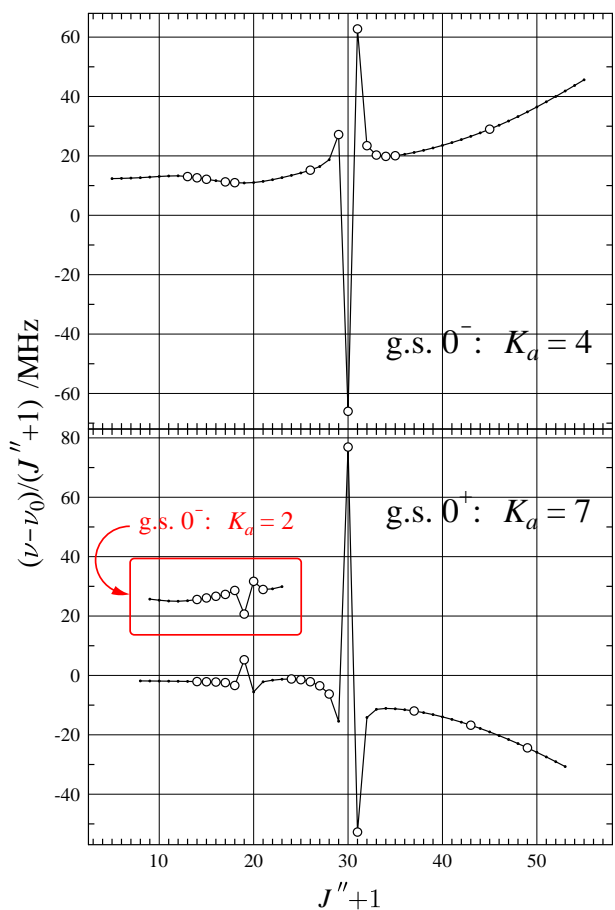}

\caption{ Perturbation shifts in selected $^aR$-branch rotational transitions in the ground state
tunneling doublet. The plotted quantity is the scaled difference
between frequencies from the full spectroscopic model ($\nu$) and perturbation
removed frequencies ($\nu_0$) calculated without the interstate coupling terms.
The matching character of the observed resonance behaviour identifies the
interacting $K_a$ sequences in the two substates. Circles denote experimental
measurements, and the perturbation contribution at the resonance peak at
$J''+1=30$ is near 2.4 GHz. The $0^+$ sequence has a smaller resonance at
$J''+1=19$ identified to result from a resonance with a different, $K_a=2$
sequence in substate $0^{\tt -}$. 
All measured transitions involved in the resonances are fitted to within their 
experimental uncertainty.
}\label{fig:pert}

\end{figure}


It is at this stage that the difficulties faced during the analysis and
possible pitfalls in searching for a new molecule in space on the basis of
incomplete fits can be appreciated. Figure~\ref{fig:pert} illustrates a typical
effect of perturbations on rotational transition frequencies in the two
substates.  Perturbations are most efficiently followed in sequences of transitions for a
given value of $K_a$ and an unperturbed situation is characterised by
smooth, near horizontal trace behaviour in a scaled frequency difference plot as in
Fig.~\ref{fig:pert}.  The sharp spikes are due to
contributions from resonances between rotational levels in the two vibrational substates, that
arise within a Hamiltonian matrix for a given value of the $J$ quantum
number.  A specific resonance between two rotational levels in perturbing
substates will have the same magnitude but the opposite sign for
the two partners.  Mirror image behaviour, such as that seen in
Fig.~\ref{fig:pert}, confirms identification of the resonance partners
and, if consistent with observed frequencies, is also indicative of the
quality of the fit.  It is notable that resonances can be of
considerable magnitude (2.4 GHz for the maxima in this figure) and can also
significantly affect transition frequencies for many $J$ values around
the maximum. There are many such resonances in the data for the
three pairs of substates, and even in Fig.~\ref{fig:pert} there is a second, smaller, resonance
in the sequence for the $0^+$ substate that has its partner in the $K_a=2$ sequence for
$0^-$.

\begin{figure}[t]
\centering
\includegraphics[width=6.0cm]{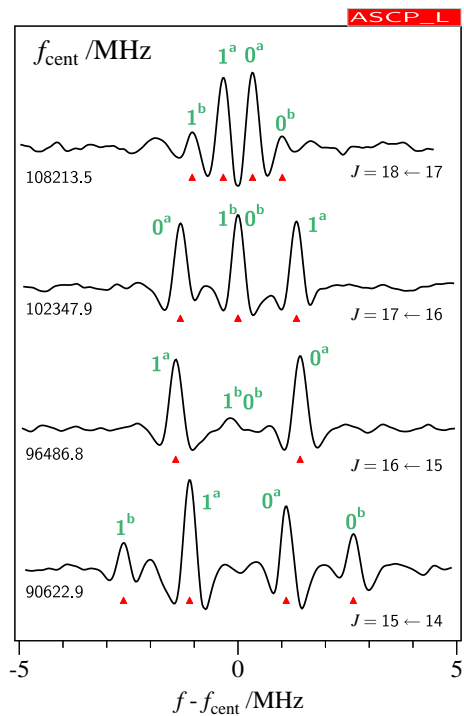}

\caption{Perturbations in the quartets of $R$-branch $K_a$=0,1 transitions in the
$0^-$ substate for the ground state of glycinamide.  At these $J$ values, the
four possible transitions are completely resolved, with the pattern of the two
stronger, central  $a$-dipole transitions, flanked by weaker $b$-dipole
transitions (as in the bottom and top traces).  The transitions are marked by
the value of $K_a$ and transition type, and the two middle traces show
significant deviations in the patterns.  Triangle markers indicate transitions
in the fitted data set.  While this is an interesting spectroscopic curiosity
it is noted that such lowest $K_a$ transitions are the first choice in astrophysical
searches for a new molecule. }\label{fig:quartets}

\end{figure}


\begin{table}[!h]
\vspace{0.2cm}
\caption{Partition function for glycinamide$^a$ and its dependence
         on accounting for vibrational states.}
\vspace{-0.5cm}
\label{tab:partf}
\renewcommand{\thefootnote}{\alph{footnote}}
\begin{center}
\begin{footnotesize}
\begin{tabular}{r rrr}
\hline\hline\vspace{-0.2cm}\\
$T$ /K                           & \multicolumn{1}{c}{ g.s.$^b$ }
                                 & \multicolumn{1}{c}{ three pairs$^c$ }
                                 & \multicolumn{1}{c}{ $\le$ 412 cm$^{-1}$~$^d$} \\
\vspace{-0.2cm}\\
\hline
\\
    300       & ~~~~160929       &     317605      &     544700      \\
    250       &     122291       &     225975      &     355078      \\
    200       &      87037       &     147233      &     207313      \\
    160       &      61777       &      95103      &     120650      \\
    150       &      55923       &      83734      &     103323      \\
    100       &      29786       &      37705      &      40729      \\
     75       &      18942       &      21739      &      22387      \\
     50       &       9901       &      10392      &      10444      \\
     25       &       3139       &       3147      &       3147      \\
\\
\hline
\end{tabular}
\end{footnotesize}
\end{center}
{\bfseries Notes.}\\
$^{(a)}$Evaluated numerically by counting all rotational states up to $J$=100.
$^{(b)}$Accounting for only the ground state $0^+$, $0^-$ doublet.
$^{(c)}$Accounting for the g.s., $v_{27}$=1, and $v_{26}$=1 tunneling doublets studied in this work.
$^{(d)}$Accounting for 19 tunneling substates estimated to be present up to 412 cm$^{-1}$ by adding predictions
from a double minimum model for the $\nu_{27}$ mode to vibrational energy levels for the $\nu_{26}$, $\nu_{25}$, and $\nu_{24}$
 normal modes, and assuming average ground state rotational constants for all vibrational states.

\vspace{0.5cm}
\end{table}

The perturbations not only affect transition frequencies, but
can also significantly affect transition intensities, as visible in
Fig.~\ref{fig:quartets}.  Without perturbations the intensity pattern
for the displayed quartet of lines is expected to be as in the bottom or
top traces.  In the present case, both the intensity pattern and quantum number
labeling of the lines in the two middle plots are significantly affected.

\begin{figure*}[!ht]
\centerline{\resizebox{0.75\hsize}{!}{\includegraphics[angle=0]{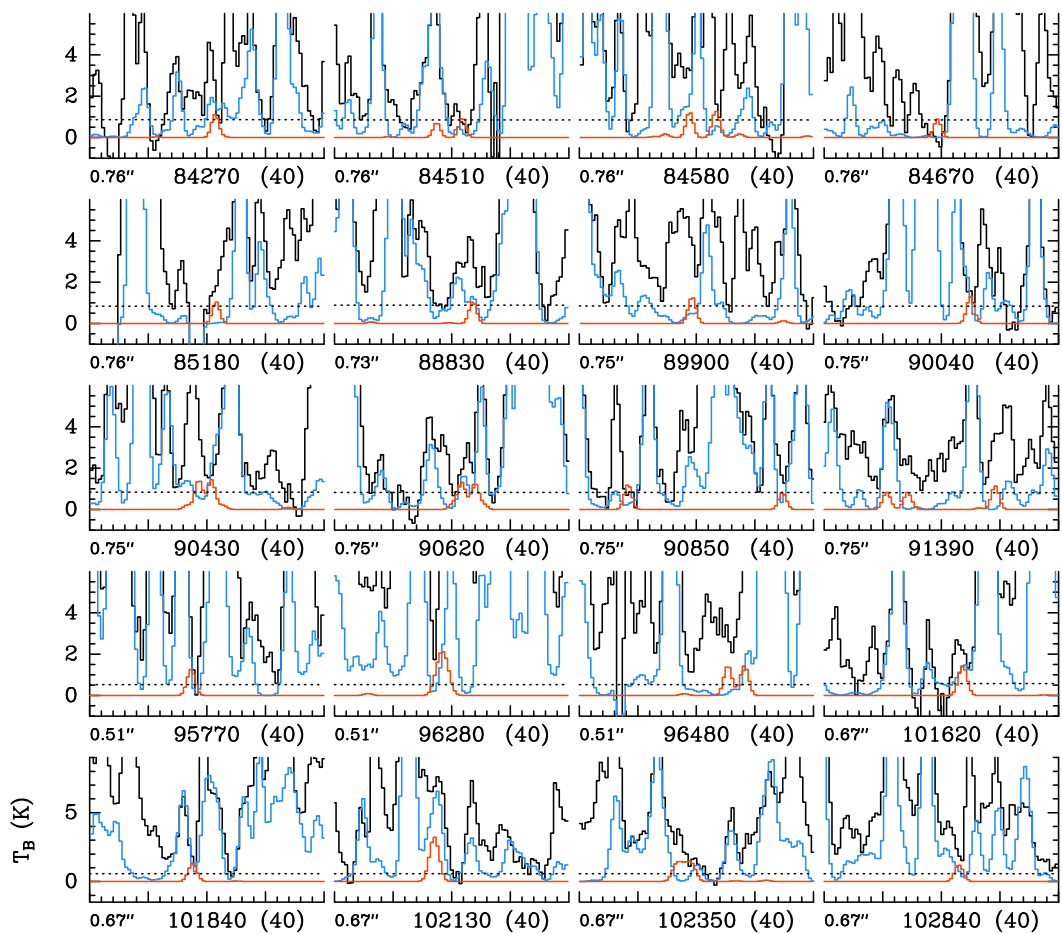}}}
\caption{Selection of transitions of NH$_2$CH$_2$C(O)NH$_2$, $\varv = 0$
covered by our ALMA survey. The synthetic spectrum of NH$_2$CH$_2$C(O)NH$_2$,
$\varv = 0$ used to derive the upper limit to its column density is displayed
in red and overlaid on the observed spectrum of Sgr~B2(N1S) shown in black.
The blue synthetic spectrum contains the contributions from all molecules
identified in our survey so far, but not from the species shown in red.
The central frequency of each panel is indicated in MHz below its 
\textit{x}-axis. Each panel has a width of 40 MHz, as indicated in 
brackets behind the central frequency.
The angular resolution (HPBW) is also indicated. The $y$-axis is labeled in
brightness temperature units (K). The dotted line indicates the $3\sigma$
noise level.}
\label{f:spec_nh2ch2conh2_ve0}
\end{figure*}

\begin{figure*}[!h]
\centerline{\resizebox{0.75\hsize}{!}{\includegraphics[angle=0]{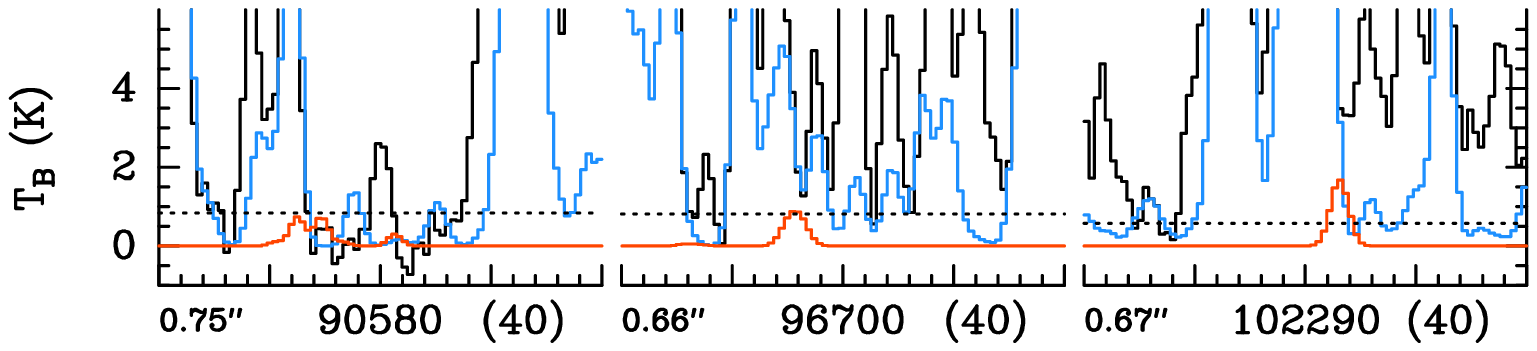}}}
\caption{Same as Fig.~\ref{f:spec_nh2ch2conh2_ve0} but for
NH$_2$CH$_2$C(O)NH$_2$, $\varv_{27} = 1$.}
\label{f:spec_nh2ch2conh2_v27e1}
\end{figure*}

\begin{figure}[!h]
\centerline{\resizebox{0.6\hsize}{!}{\includegraphics[angle=0]{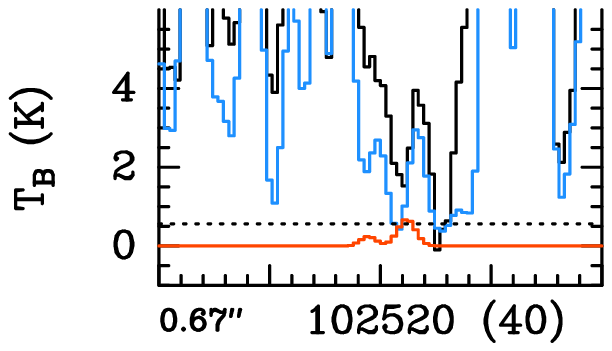}}}
\caption{Same as Fig.~\ref{f:spec_nh2ch2conh2_ve0} but for
NH$_2$CH$_2$C(O)NH$_2$, $\varv_{26} = 1$.}
\label{f:spec_nh2ch2conh2_v26e1}
\end{figure}

Finally the number of low lying vibrational states has a significant
effect on the value of the vibration-rotation partition function $Q_{vr}$ that is of
relevance to column density evaluations.  The vibrational effect for glycinamide is documented in
Table~\ref{tab:partf}.  It is seen that at the lowest temperatures the
often used ground state partition function is quite satisfactory.  Yet
at 200~K, for example, the partition function correction arising from the presence of also the $v_{27}$=1
and $v_{26}$=1 doublets is 1.7.  Glycinamide also has
two other relatively low frequency modes ($\nu_{25}$ and $\nu_{24}$ calculated at
303 and 384~cm$^{-1}$ resp., \citealt{Li2003}).  
Combination of those modes with present
information on $\nu_{27}$ and $\nu_{26}$ estimates the need to account for populating
a total of 19 sublevels up to 412 cm$^{-1}$ corresponding to a partition function correction of 2.4
at 200~K.

%
\section{Search for glycinamide toward Sgr~B2(N1)}
\label{s:astro}

\subsection{Observations}
\label{ss:observations}

We used the imaging spectral line survey ReMoCA (Re-exploring Molecular Complexity with ALMA)
performed with ALMA toward
Sgr~B2(N). The observational setup and the method used to reduce this
interferometric data set were described in \citet{Belloche19}. We summarise
only the main features here. The field of view of the observations was centered
between Sgr~B2(N1) and Sgr~B2(N2), the two main hot molecular cores of
Sgr~B2(N), which are separated by 4.9$\arcsec$ or $\sim$0.2~pc in projection
onto the plane of the sky. The equatorial coordinates of this phase center are
($\alpha, \delta$)$_{\rm J2000}$=
($17^{\rm h}47^{\rm m}19{\fs}87, -28^\circ22'16{\farcs}0$). The survey has an
angular resolution (HPBW) that varies between $\sim$0.3$\arcsec$ and
$\sim$0.8$\arcsec$. The median angular resolution is 0.6$\arcsec$ and
corresponds to $\sim$4900~au at the distance of Sgr~B2
\citep[8.2~kpc,][]{Reid19}. Five frequency tunings of the receivers, which we
call setups S1 to S5, were used to cover the frequency range from 84.1~GHz to
114.4~GHz at a spectral resolution of 488~kHz (1.7 to 1.3~km~s$^{-1}$). The
observations achieved a sensitivity per spectral channel between
0.35~mJy~beam$^{-1}$ and 1.1~mJy~beam$^{-1}$ (rms) depending on the setup, with
a median sensitivity of 0.8~mJy~beam$^{-1}$.

Following \citet{Belloche19} we selected the offset position Sgr~B2(N1S)
located at ($\alpha, \delta$)$_{\rm J2000}$=
($17^{\rm h}47^{\rm m}19{\fs}870$, $-28^\circ22\arcmin19{\farcs}48$) for this
study. This position is about 1$\arcsec$ to the south of the main hot core
Sgr~B2(N1). Its continuum emission has a lower opacity than the peak of the
hot core, which allows for a deeper look into the molecular content of this
source. We used here an improved version of the data reduction, as reported in
\citet{Melosso20}.

We employed the software Weeds \citep[][]{Maret11} to produce synthetic spectra
under the assumption of local thermodynamic equilibrium (LTE). This assumption
is appropriate for Sgr~B2(N1S) because the regions where hot-core emission is
detected in Sgr~B2(N) have high densities
\citep[$>1 \times 10^{7}$~cm$^{-3}$, see][]{Bonfand19}. A best-fit
synthetic spectrum was derived for each molecule separately, and then the
contributions of all identified molecules were added together. Each
species was modeled with a set of five parameters: size of the emitting region
($\theta_{\rm s}$), column density ($N$), temperature ($T_{\rm rot}$), linewidth
($\Delta V$), and velocity offset ($V_{\rm off}$) with respect to the assumed
systemic velocity of the source, $V_{\rm sys}=62$~km~s$^{-1}$.

\begin{table*}[!ht]
 \begin{center}
 \caption{
 Parameters of our best-fit LTE models of cyanamide, aminoacetonitrile, formamide, methylamine, and urea toward Sgr~B2(N1S), 
 along with column density upper limit for glycinamide.
}
 \label{t:coldens}
 \vspace*{-1.2ex}
 \begin{tabular}{lcrcccccc}
 \hline\hline
 \multicolumn{1}{l}{Molecule} & \multicolumn{1}{c}{Status\tablefootmark{a}} & \multicolumn{1}{c}{$N_{\rm det}$\tablefootmark{b}} & \multicolumn{1}{c}{$\theta_{\rm s}$\tablefootmark{c}} & \multicolumn{1}{c}{$T_{\mathrm{rot}}$\tablefootmark{d}} & \multicolumn{1}{c}{$N$\tablefootmark{e}} & \multicolumn{1}{c}{$F_{\rm vib}$\tablefootmark{f}} & \multicolumn{1}{c}{$\Delta V$\tablefootmark{g}} & \multicolumn{1}{c}{$V_{\mathrm{off}}$\tablefootmark{h}} \\ 
 \multicolumn{1}{c}{     } & & & \multicolumn{1}{c}{\small ($''$)} & \multicolumn{1}{c}{\small (K)} & \multicolumn{1}{c}{\small (cm$^{-2}$)} & & \multicolumn{1}{c}{\small (km~s$^{-1}$)} & \multicolumn{1}{c}{\small (km~s$^{-1}$)} \\ 
 \hline
 NH$_2$CN & d & 3 &  2.0 &  160 &  2.6 (16) & 1.03 & 5.5 & -0.3 \\ 
 NH$_2$CH$_2$CN\tablefootmark{i} & d & 23 &  2.0 &  200 &  1.1 (17) & 1.00 & 5.0 & 0.0 \\ 
 NH$_2$CHO\tablefootmark{j} & d & 34 &  2.0 &  160 &  2.9 (18) & 1.09 & 6.0 & 0.0 \\ 
 NH$_2$CH$_3$ & d & 15 &  2.0 &  230 &  1.4 (18) & 1.25 & 5.0 & 0.0 \\ 
 NH$_2$C(O)NH$_2$\tablefootmark{j} & d & 9 &  2.0 &  160 &  2.7 (16) & 1.86 & 5.0 & 0.0 \\ 
 NH$_2$CH$_2$C(O)NH$_2$ & n & 0 &  2.0 &  160 & $<$  1.5 (16) & 1.27 & 5.0 & 0.0 \\ 
\hline 
 \end{tabular}
 \end{center}
 \vspace*{-2.5ex}
 \tablefoot{
 \tablefoottext{a}{d: detection, n: nondetection.}
 \tablefoottext{b}{Number of detected lines \citep[conservative estimate, see Sect.~3 of][]{Belloche16}. One line of a given species may mean a group of transitions of that species that are blended together.}
 \tablefoottext{c}{Source diameter (\textit{FWHM}).}
 \tablefoottext{d}{Rotational temperature.}
 \tablefoottext{e}{Total column density of the molecule. $x$ ($y$) means $x \times 10^y$.}
 \tablefoottext{f}{Correction factor that was applied to the column density to account for the contribution of vibrationally excited states, in the cases where this contribution was not included in the partition function of the spectroscopic predictions.}
 \tablefoottext{g}{Linewidth (\textit{FWHM}).}
 \tablefoottext{h}{Velocity offset with respect to the assumed systemic velocity of Sgr~B2(N1S), $V_{\mathrm{sys}} = 62$ km~s$^{-1}$.}
 \tablefoottext{i}{The parameters were derived from the ReMoCA survey by \citet{Melosso20}.}
 \tablefoottext{j}{The parameters were derived from the ReMoCA survey by \citet{Belloche19}.}
 }
 \end{table*}

\subsection{Nondetection of glycinamide}
\label{ss:nondetection}

We assumed a rotational temperature of 160~K, as derived for formamide,
NH$_2$CHO, by \citet{Belloche19}, an emission size of 2$\arcsec$, and an FWHM
linewidth of 5~km~s$^{-1}$ to compute LTE synthetic spectra of glycinamide
and search for rotational emission of this molecule toward Sgr~B2(N1S). We
found no evidence for emission of glycinamide toward this source. This
nondetection is illustrated in Fig.~\ref{f:spec_nh2ch2conh2_ve0}. We also
searched for rotational emission from within its vibrationally excited states
$\varv_{27}=1$ and $\varv_{26}=1$ but did not detect any clear sign of it (see
Figs.~\ref{f:spec_nh2ch2conh2_v27e1} and \ref{f:spec_nh2ch2conh2_v26e1},
respectively). The upper limit that we obtain for the column density of
glycinamide is indicated in
Table~\ref{t:coldens} along with the parameters derived by \citet{Belloche19}
for urea, NH$_2$C(O)NH$_2$, and formamide, NH$_2$CHO, and by \citet{Melosso20}
for aminoacetonitrile, NH$_2$CH$_2$CN.

The amides NH$_2$CH$_2$C(O)NH$_2$ and NH$_2$C(O)NH$_2$ could be seen as the
partially hydrolyzed counterparts of the nitriles NH$_2$CH$_2$CN and NH$_2$CN,
respectively. It may thus be instructive to compare the relative abundances of
these two pairs of molecules. With this in mind, we modeled the emission
spectrum of cyanamide (aminomethanenitrile, NH$_2$CN) toward Sgr~B2(N1S). 
The cyanamide spectroscopic information was taken from the JPL catalog 
\citep{jplcatalog}. This JPL entry (TAG 42003 version 1) is based 
mostly on microwave data from \citet{Read1986} along with far-infrared 
data from \citet{Birk1993}.
A few transitions of cyanamide are clearly detected (see
Fig.~\ref{f:spec_nh2cn_ve0}), but unfortunately not enough to derive its
rotational temperature from a population diagram. Therefore, we also assumed a
temperature of 160~K to compute its column density, which is given in
Table~\ref{t:coldens}. The cyanamide line around 100.070~GHz is contaminated by
absorption from HC$_3$N $J$=11--10, which is not accurately accounted for by
our current complete model of Sgr~B2(N1S).

\begin{table}
 \begin{center}
 \caption{
 Rotational temperature of methylamine derived from its population diagram toward Sgr~B2(N1S).
}
 \label{t:popfit}
 \vspace*{0.0ex}
 \begin{tabular}{lll}
 \hline\hline
 \multicolumn{1}{c}{Molecule} & \multicolumn{1}{c}{States\tablefootmark{a}} & \multicolumn{1}{c}{$T_{\rm fit}$\tablefootmark{b}} \\ 
  & & \multicolumn{1}{c}{\small (K)} \\ 
 \hline
NH$_2$CH$_3$ & $\varv=0$ & 223.0 (5.9) \\ 
\hline 
 \end{tabular}
 \end{center}
 \vspace*{-2.5ex}
 \tablefoot{
 \tablefoottext{a}{Vibrational states that were taken into account to fit the population diagram.}
 \tablefoottext{b}{The standard deviation of the fit is given in parentheses. As explained in Sect.~3 of \citet{Belloche16} and in Sect.~4.4 of \citet{Belloche19}, this uncertainty is purely statistical and should be viewed with caution. It may be underestimated.}
 }
 \end{table}

Radicals derived from methylamine, NH$_2$CH$_3$, may be involved in the
formation of glycinamide. Therefore, we also report in Table~\ref{t:coldens}
the parameters we derived by modelling the emission spectrum of this molecule.
The methylamine spectroscopic information was taken from the JPL 
catalog. This catalog entry (TAG 31008 version 1) is based on the 
combined fit by \citet{Ilyushin2005} with data in the range of our 
survey mostly from that work.
A dozen transitions of NH$_2$CH$_3$ are clearly detected toward
Sgr~B2(N1S), as shown in Fig.~\ref{f:spec_nh2ch3_ve0}. There are enough
transitions to build a population diagram, which is displayed in
Fig.~\ref{f:popdiag_nh2ch3}. We assumed a temperature of 230~K for the LTE
modelling of the spectrum of NH$_2$CH$_3$ shown in
Fig.~\ref{f:spec_nh2ch3_ve0}, slightly higher than the formal result of the
fit to the population diagram (see Table~\ref{t:popfit}), but within the uncertainties.

The vibrational correction factor $F_{\rm vib}$ reported in 
Table~\ref{t:coldens} for glycinamide corresponds to the ratio of the values 
given in columns 4 and 3 of Table~\ref{tab:partf}. This is because the 
partition function of the spectroscopic predictions used for the astronomical 
search included only the vibrational contribution given in column 3 of that
table. In proceeding like this, we neglect the contribution of vibrational 
levels above 412~cm$^{-1}$, which is likely below a further 10\% at a 
temperature of 160~K.

\section{Discussion}
\label{s:discussion}

\subsection{Laboratory spectroscopy of glycinamide}
\label{ss:labspec} 

The present experimental investigation of the rotational spectrum covers 
the most useful ALMA bands for a molecule of this size.  Many resonances 
between rotational levels in the three studied tunneling doublets have 
been satisfactorily fitted, allowing reliable predictions, at least when these are interpolations within
the acquired data sets.  At the laboratory measurement temperature of 50$^\circ$C the strongest 
observed lines have all been accounted for, providing a comprehensive basis for any future searches,
especially at the typically significantly lower interstellar temperatures.
Clearly many more outstanding lines from higher vibrational states remain to be assigned and 
measured in the experimental spectrum, but their analysis is expected to 
be even more challenging than that reported presently.

\subsection{Comparison of glycinamide to related molecules in Sgr~B2(N)}
\label{ss:comparison} 

The nondetection of glycinamide, NH$_2$CH$_2$C(O)NH$_2$, toward Sgr~B2(N1S)
reported in Table~\ref{t:coldens} implies that it is at least $\sim$1.8~times
less abundant than urea, NH$_2$C(O)NH$_2$, which is markedly different from
the pair of nitriles NH$_2$CH$_2$CN/NH$_2$CN for which the longer molecule is
a factor four more abundant than the shorter one. Said differently,
NH$_2$C(O)NH$_2$, the partially hydrolyzed counterpart of NH$_2$CN, has the
same abundance as the latter, while NH$_2$CH$_2$C(O)NH$_2$, the partially
hydrolyzed counterpart of NH$_2$CH$_2$CN, is at least seven times less
abundant than the latter. The analysis reported in Sect.~\ref{s:astro} also
shows that glycinamide is at least two orders of magnitude less abudant than
its potential precursors formamide and methylamine.

\subsection{Glycinamide chemistry}
\label{ss:chemistry}

Glycinamide is not presently included in any astrochemical models, but 
the results of recent chemical simulations of hot cores may nevertheless be 
informative. \citet{Garrod2013} constructed a chemical network for glycine 
(NH$_2$CH$_2$COOH), which included the related species glycinal 
(NH$_2$CH$_2$CHO). These species could be formed on interstellar dust grains 
through the addition of radicals produced mainly by photodissocation of 
simpler solid-phase molecules, driven by the enhanced surface mobility of the 
radicals, as the result of grain heating induced by the star-formation 
process. Complex organic molecules including glycine and glycinal would later 
sublimate entirely from the dust grains. The chemical network included 
mechanisms for those molecules to be subsequently destroyed in the gas phase 
through ion-molecule reactions.

The more recent models by \citet{Garrod21}, which incorporate the same 
glycine-related chemistry, allow a broader range of grain-surface and bulk-ice 
kinetic processes to bring together reactive radicals and thus form complex 
organics. This includes the possibility of radicals reacting nondiffusively on 
dust-grain surfaces at very low temperatures, when the icy grain mantles are 
still gradually building up. In this scenario, the radicals themselves need 
only be formed close to each other, allowing their reactions to proceed 
unmediated by thermal diffusion. Furthermore, the initially translucent 
conditions under which the dust-grain ices build up may allow some degree of 
UV photoprocessing of the young ices by the ambient interstellar UV field, 
converting a small fraction of simple solid-phase molecules into more complex organics. 
The \citeauthor{Garrod21} models indicate that a substantial fraction of 
glycine production may occur at this very early stage, with glycinal acting as 
a precursor. Although it is not included in those models, glycinamide could 
plausibly be produced through a similar mechanism; i.e., the photodissociation 
of solid-phase glycinal to produce the radical NH$_2$CH$_2$CO, with which 
NH$_2$ would react to form NH$_2$CH$_2$CONH$_2$.

If such a mechanism is active in producing solid-phase glycinamide, 
then the observed ratio of formamide to urea might be somewhat indicative of 
the expected ratio of glycinal to glycinamide, which share similar molecular 
structures. Toward Sgr~B2(N1S), the former ratio is approximately 100. Although 
glycinal is thus far undetected in the interstellar medium, its search being
hindered by a lack of spectroscopic predictions, the model of 
\citet{Garrod21} suggests that its peak gas phase abundance should be around a 
factor 100 less than that of formamide. On this basis, one might therefore 
expect an abundance of glycinamide around 10$^4$ times lower than formamide, 
or 100 times lower than urea.
However, it is plausible that the efficiency of the conversion of 
glycinal to glycinamide on the grains could be greater than the above 
crude comparison might suggest; in the models, glycine itself reaches a 
peak gas-phase abundance as great as around half that of glycinal, 
implying substantial conversion. This corresponds to an abundance 
approximately 40 times lower than that of urea produced by the model. If 
we were to take this abundance of glycine as the maximum possible 
allowed abundance for glycinamide, which assumes that glycine and 
glycinamide would be produced in equal amounts from glycinal, then the 
implied 40:1 ratio of urea to glycinamide would produce a glycinamide 
column density around 20 times lower than the observed upper limit. Even 
efficient conversion from glycinal would therefore still produce a 
rather smaller amount of glycinamide than might plausibly be detectable 
at this time.

The above production mechanism is, however, not the only possible 
means by which glycinamide could form through radical addition; alternatives 
would include the reaction of the acetamide-related radical NH$_2$COCH$_2$ 
with NH$_2$, or the addition of NH$_2$CH$_2$ to NH$_2$CO. The latter pair of 
radicals can be formed through cosmic ray-induced UV photodissociation of, or 
chemical H-atom abstraction from, solid-phase NH$_2$CH$_3$ and NH$_2$CHO, both 
of which are detected toward Sgr B2(N1S) in the gas phase. A similar reaction 
between radicals NH$_2$CH$_2$ and COOH indeed produces a substantial quantity 
of glycine in the models; this process occurs much later, at around the time 
when water itself is beginning to desorb rapidly from the grains, releasing 
trapped radicals onto the surface of the warm ice, where they may react 
diffusively. The sizeable abundances of NH$_2$CH$_3$ and NH$_2$CHO detected 
toward Sgr B2(N1S) suggest this could be a plausible pathway for glycinamide 
production.

The above reaction mechanisms nevertheless remain entirely 
conjectural, at least until they have been tested explicitly in the 
astrochemical models. The closure of the existing chemical network surrounding 
glycine, to incorporate glycinamide, seems a plausible goal for future 
investigation.

\section{Conclusions}
A comprehensive laboratory rotational study of glycinamide, a potential glycine precursor,
has been undertaken up to 329 GHz. In total, over 2800 transition lines were assigned
and measured for the ground state and two lowest-lying excited state tunneling doublets.
Newly derived spectroscopic constants were used to search for spectral signatures of glycinamide
in Sgr~B2(N) by millimetre-wave astronomy.  Lists of experimental frequencies and their obs.-calc. 
differences are provided in Tables 5, 6, and 7, only available in electronic form.

Glycinamide was not detected toward the hot molecular core Sgr~B2(N1S)
with ALMA. The upper limit derived for its column density implies that it is at
least seven times less abundant than aminoacetonitrile and 1.8 times less
abundant than urea toward this source.

While glycinamide has not yet been considered in any astrochemical kinetics models, 
comparison with model results for
related species suggest that it may be a factor of 40 to
100 times less abundant than urea, corresponding to a value a factor of 20 to 50 below the
upper limit toward Sgr B2(N1S).
This would likely be well below the spectral confusion limit of the
ReMoCA survey of Sgr B2(N). Further progress in the search for
glycinamide in the interstellar medium could be made by targetting 
sources with a lower level of spectral confusion. The Galactic
Center source G+0.693-0.027, a shocked region close to Sgr B2(N)
with a rich chemical content characterized by low excitation 
temperatures, should be a promising target to continue the search
for interstellar glycinamide.

\begin{acknowledgements} This paper makes use of the following ALMA 
data: ADS/JAO.ALMA\#2016.1.00074.S. ALMA is a partnership of ESO 
(representing its member states), NSF (USA), and NINS (Japan), together 
with NRC (Canada), NSC and ASIAA (Taiwan), and KASI (Republic of Korea), 
in cooperation with the Republic of Chile. The Joint ALMA Observatory is 
operated by ESO, AUI/NRAO, and NAOJ. The interferometric data are 
available in the ALMA archive at https://almascience.eso.org/aq/. Part 
of this work has been carried out within the Collaborative Research 
Centre 956, sub-project B3, funded by the Deutsche 
Forschungsgemeinschaft (DFG) -- project ID 184018867. 
JCG thanks the National Center for Space Studies 
(CNES), National Program “Physics and Chemistry of the Interstellar 
Medium” (PCMI) from CNRS / INSU, and PHC Polonium 14277ZC for financial 
support. LK and Valladolid authors acknowledge funding from the Ministerio de 
Ciencia e Innovación (grants PID2019-111396GB-I00, CTQ2016-76393-P), 
Junta de Castilla y León (grants VA244P20, VA077U16), and European 
Research Council under the European Union’s Seventh Framework Programme 
ERC-2013- SyG, Grant Agreement n. 610256 NANOCOSMOS. 
RTG acknowledges support from the National Science Foundation (grant No. AST 19-06489).
ZK and EBJ also acknowledge 
the use of computational resources under the grant G61-6 from the 
Interdisciplinary Center of Mathematical and Computer Modelling (ICM) of Warsaw University.

\end{acknowledgements}



\bibliography{library}

\clearpage

\begin{appendix}
\label{appendix}

\section{Complementary Tables and figures}
\label{a:spectra}

Table~\ref{tab:subsets} contains subset statistics for substates in each 
torsional/inversion doublet, including numbers of fitted lines, ranges 
of the values of the key quantum numbers, and frequency ranges of the 
measurements.

\vspace{0.5cm}

\begin{figure*}[!h]
\vspace{1.0cm}
\centerline{\resizebox{1.0\hsize}{!}{\includegraphics[angle=-90]{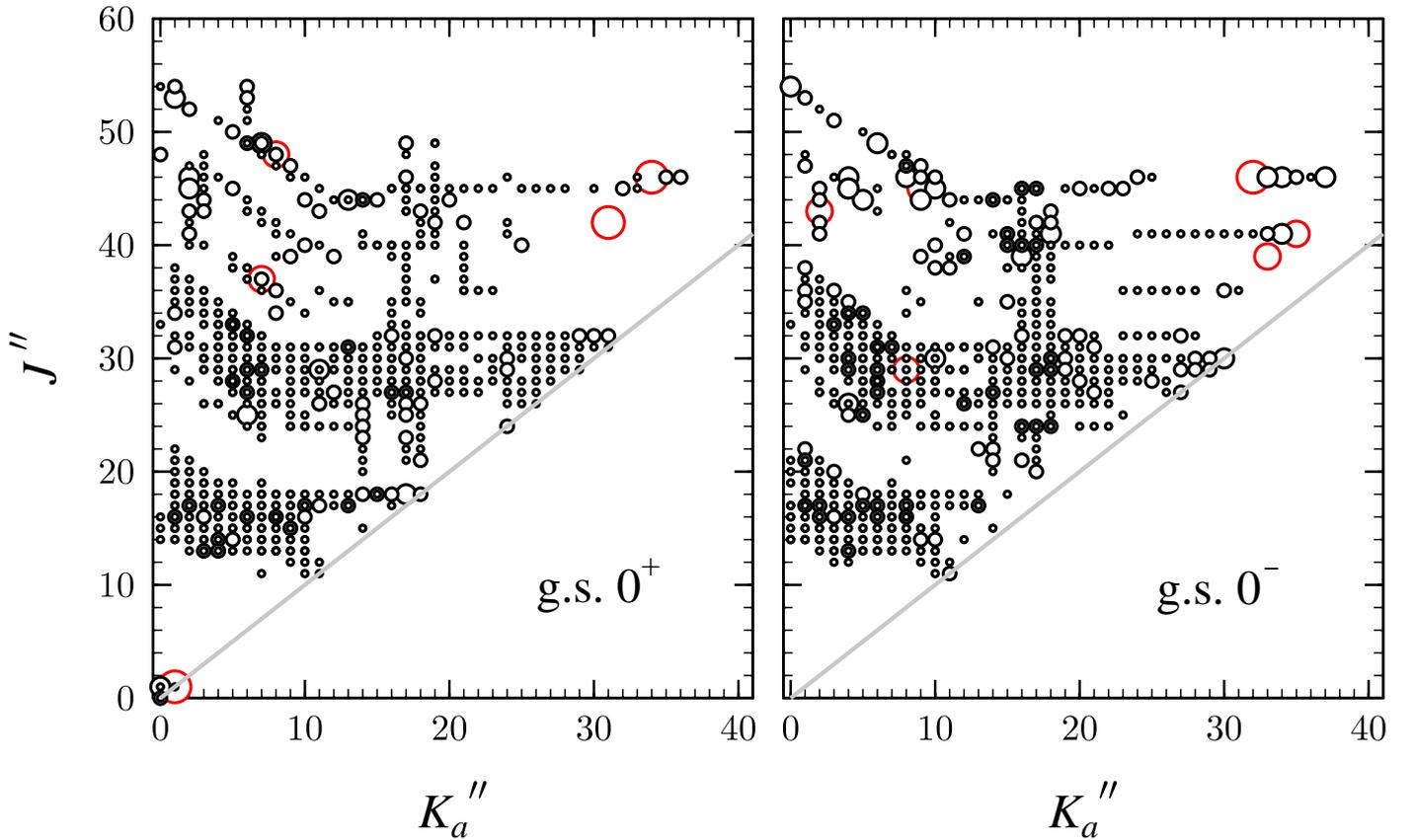}}}
\caption{Distribution plot of quantum numbers of rotational transitions measured and fitted
for the ground state doublet of glycinamide.  Symbol diameter is proportional to the value of
$|f_{\rm obs-calc}|/\sigma$ where $f_{\rm obs-calc}$ is the residual of fit for a given line
and $\sigma$ is its measurement uncertainty.  Red colour identifies
outliers with $|f_{\rm obs-calc}|/\sigma > 3$.
The few outliers are all for confidently assigned transitions and may be due either to blends with
unassigned lines or incompletely treated perturbation contributions.
}
\label{fig:distr0}
\end{figure*}

\begin{figure*}[!h]
\centerline{\resizebox{0.9\hsize}{!}{\includegraphics[angle=-90]{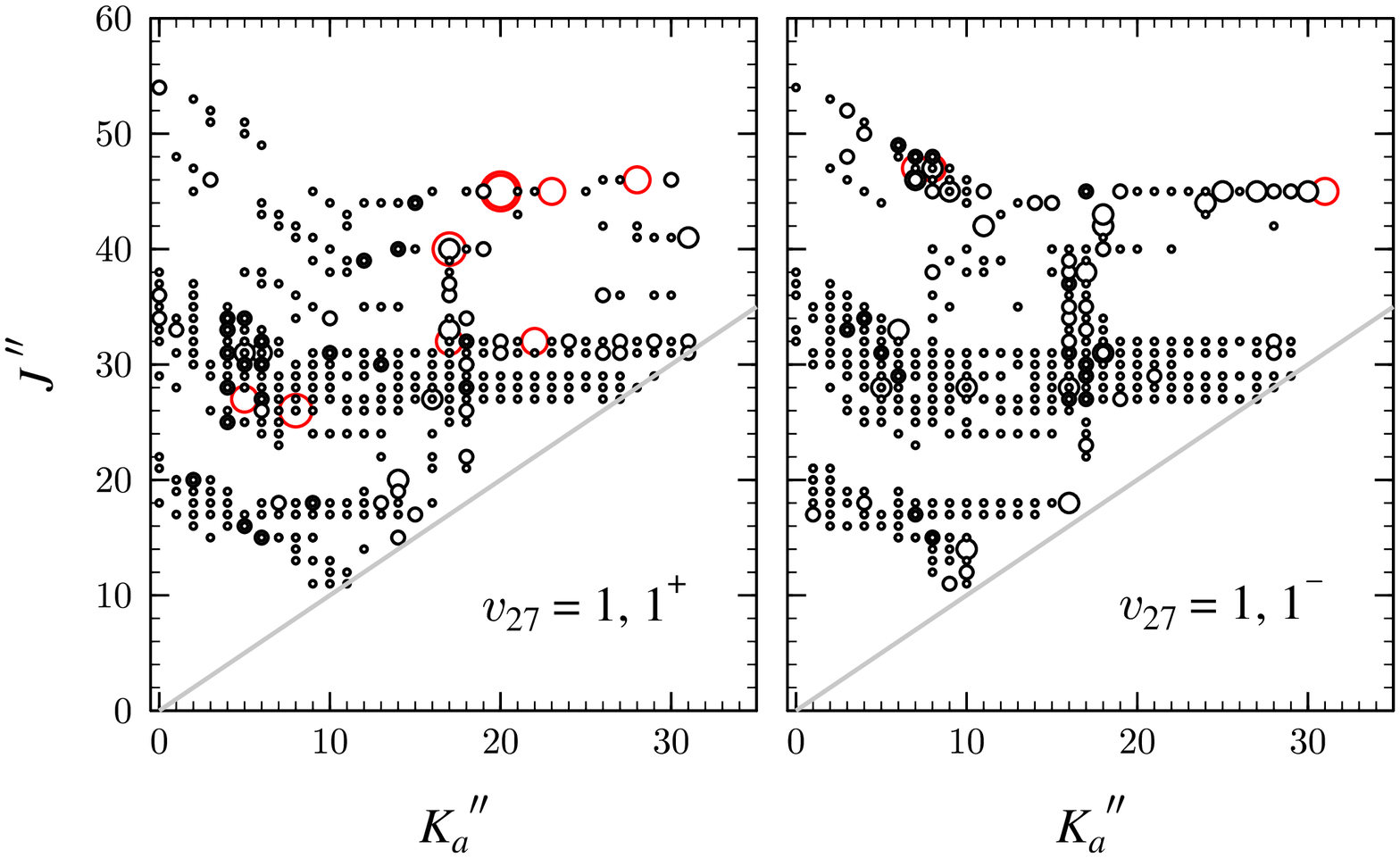}}}
\caption{Same as Fig.~\ref{fig:distr0} but for 
the $\varv_{27} = 1$ doublet of glycinamide.
}
\label{fig:distr1}
\vspace{1cm}
\end{figure*}

\begin{figure*}[!h]
\centerline{\resizebox{0.9\hsize}{!}{\includegraphics[angle=-90]{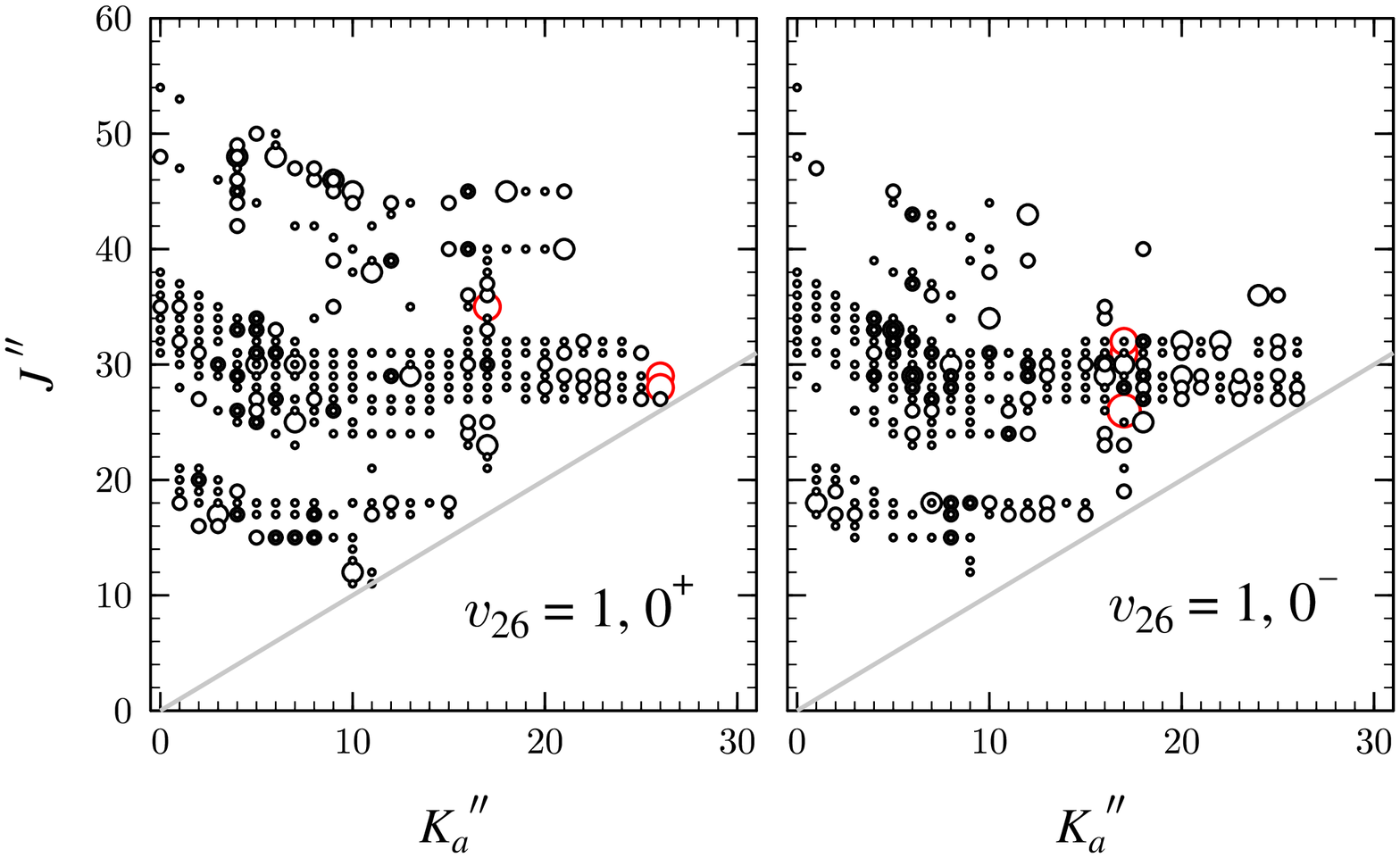}}}
\caption{Same as Fig.~\ref{fig:distr0} but for 
the $\varv_{26} = 1$ doublet of glycinamide.}
\label{fig:distr2}
\end{figure*}

The plots in Figs.~\ref{fig:distr0}, ~\ref{fig:distr1}, 
~\ref{fig:distr2} are distribution plots of obs.-calc. frequencies as a 
function of the values of $J"$ and $K_a"$ quantum numbers, illustrating 
the comprehensive quantum number coverage achieved in the measurements.

\vspace{0.5cm}

\begin{figure*}[]
\centerline{\resizebox{0.88\hsize}{!}{\includegraphics[angle=0]{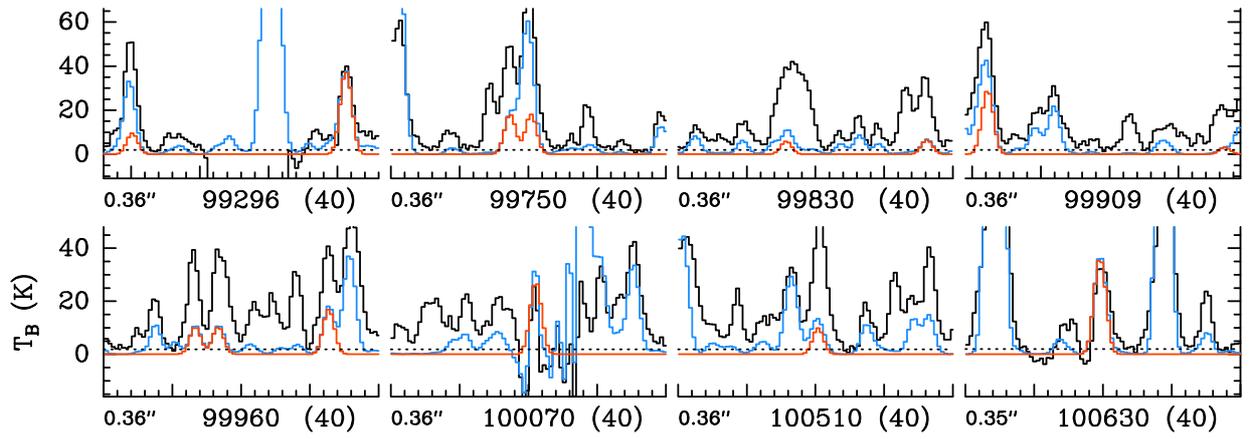}}}
\caption{Transitions of NH$_2$CN, $\varv = 0$ covered by our ALMA survey. The
synthetic spectrum of NH$_2$CN, $\varv = 0$ is displayed in red and overlaid
on the observed spectrum of Sgr~B2(N1S) shown in black. The blue synthetic
spectrum contains the contributions from all molecules identified in our
survey so far, including the species shown in red.
The central frequency of each panel is indicated in MHz below its 
\textit{x}-axis. Each panel has a width of 40 MHz, as indicated in 
brackets behind the central frequency.
The angular resolution (HPBW) is also indicated. 
The $y$-axis is labeled in
brightness temperature units (K). The dotted line indicates the $3\sigma$
noise level.}
\label{f:spec_nh2cn_ve0}
\end{figure*}

Figures~\ref{f:spec_nh2cn_ve0} and \ref{f:spec_nh2ch3_ve0} show the
transitions of NH$_2$CN and NH$_2$CH$_3$ that are covered by the ReMoCA survey
and contribute significantly to the signal detected toward Sgr~B2(N1S).
Figure~\ref{f:popdiag_nh2ch3} shows the population diagram of NH$_2$CH$_3$
toward Sgr~B2(N1S).

\vspace{0.5cm}

\begin{table*}[!h]
\vspace{ 0.5cm}
\begin{center}
{
\caption{Subset statistics for the coupled state fits of the three tunneling doublets for glycinamide.
}
\vspace{0.5cm}
\label{tab:subsets}
\renewcommand{\thefootnote}{\alph{footnote}}
\begin{small}
\begin{tabular}{lr  k{4}k{4} k{2}k{2}  ck{2}k{2} ck{5}k{5} }

\hline\vspace{-0.2cm}\\
Substate &\multicolumn{1}{c}{$N_{\rm lines}$$^a$}   &
         \multicolumn{1}{c}{$\sigma^b$/MHz}          &
         \multicolumn{1}{c}{${\sigma_{\rm rms}}^c$}    &
         \multicolumn{2}{c}{$J$}                   &&
         \multicolumn{2}{c}{$K_a$}                 &&
         \multicolumn{2}{c}{frequency /GHz} \\

        \cline{5-6} \cline{8-9} \cline{11-12}
       &&&&
       \multicolumn{1}{c}{min} &
       \multicolumn{1}{c}{max} &&
       \multicolumn{1}{c}{min} &
       \multicolumn{1}{c}{max} &&
       \multicolumn{1}{c}{min} &
       \multicolumn{1}{c}{max} \\
\vspace{-0.3cm}\\
\hline
%
                          &         &           &           &       &       &&      &      &&           &             \\
ground state, 0$^+$       & 635$^d$ &  0.0405   &  0.8353   &  0    &  55   &&   0  & 36   &&    6.706  &  328.902    \\
ground state, 0$^-$       & 602$^e$ &  0.0469   &  0.9354   & 11    &  55   &&   0  & 37   &&   90.620  &  328.918    \\
                          &         &           &           &       &       &&      &      &&           &             \\
$\varv_{27}$, 1$^+$       & 438     &  0.0457   &  0.9129   & 11    &  55   &&   0  & 31   &&  112.147  &  328.889    \\
$\varv_{27}$, 1$^-$       & 435     &  0.0429   &  0.8578   & 11    &  55   &&   0  & 31   &&  112.210  &  328.346    \\
                          &         &           &           &       &       &&      &      &&           &             \\
$\varv_{26}$, 0$^+$       & 381     &  0.0491   &  0.9822   & 11    &  55   &&   0  & 26   &&  112.148  &  328.787    \\
$\varv_{26}$, 0$^-$       & 333     &  0.0516   &  1.0316   & 12    &  55   &&   0  & 26   &&  112.238  &  326.637    \\
                          &         &           &           &       &       &&      &      &&           &             \\
\hline\vspace{-0.2cm}\\
\end{tabular}
\end{small}
}
\end{center}
{\bfseries Notes.}\\
$^{(a)}${Number of distinct measured frequencies.}
$^{(b)}$Standard deviation of the fit.
$^{(c)}$Unitless (weighted) deviation of the fit.
$^{(d)}$Including 18 nominal $0^- \leftarrow 0^+$ transitions.
$^{(e)}$Including 15 nominal $0^+ \leftarrow 0^-$ transitions.

\vspace{1cm}
\end{table*}

\vspace{0.5cm}


\begin{figure*}[!t]
\centerline{\resizebox{0.88\hsize}{!}{\includegraphics[angle=0]{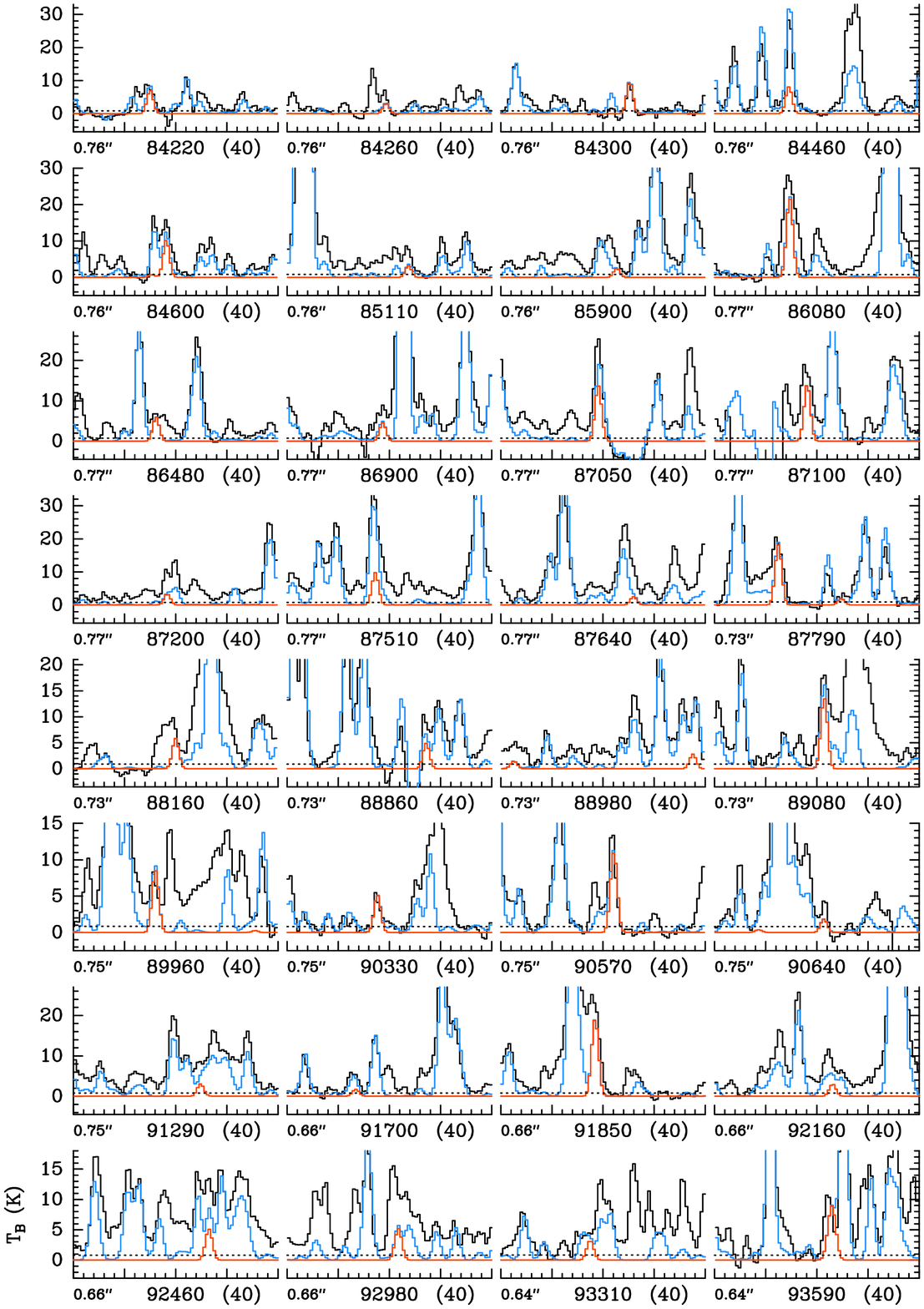}}}
\caption{Same as Fig.~\ref{f:spec_nh2cn_ve0} but for NH$_2$CH$_3$, $\varv = 0$.
}
\label{f:spec_nh2ch3_ve0}
\end{figure*}

\begin{figure*}
\addtocounter{figure}{-1}
\centerline{\resizebox{0.9\hsize}{!}{\includegraphics[angle=0]{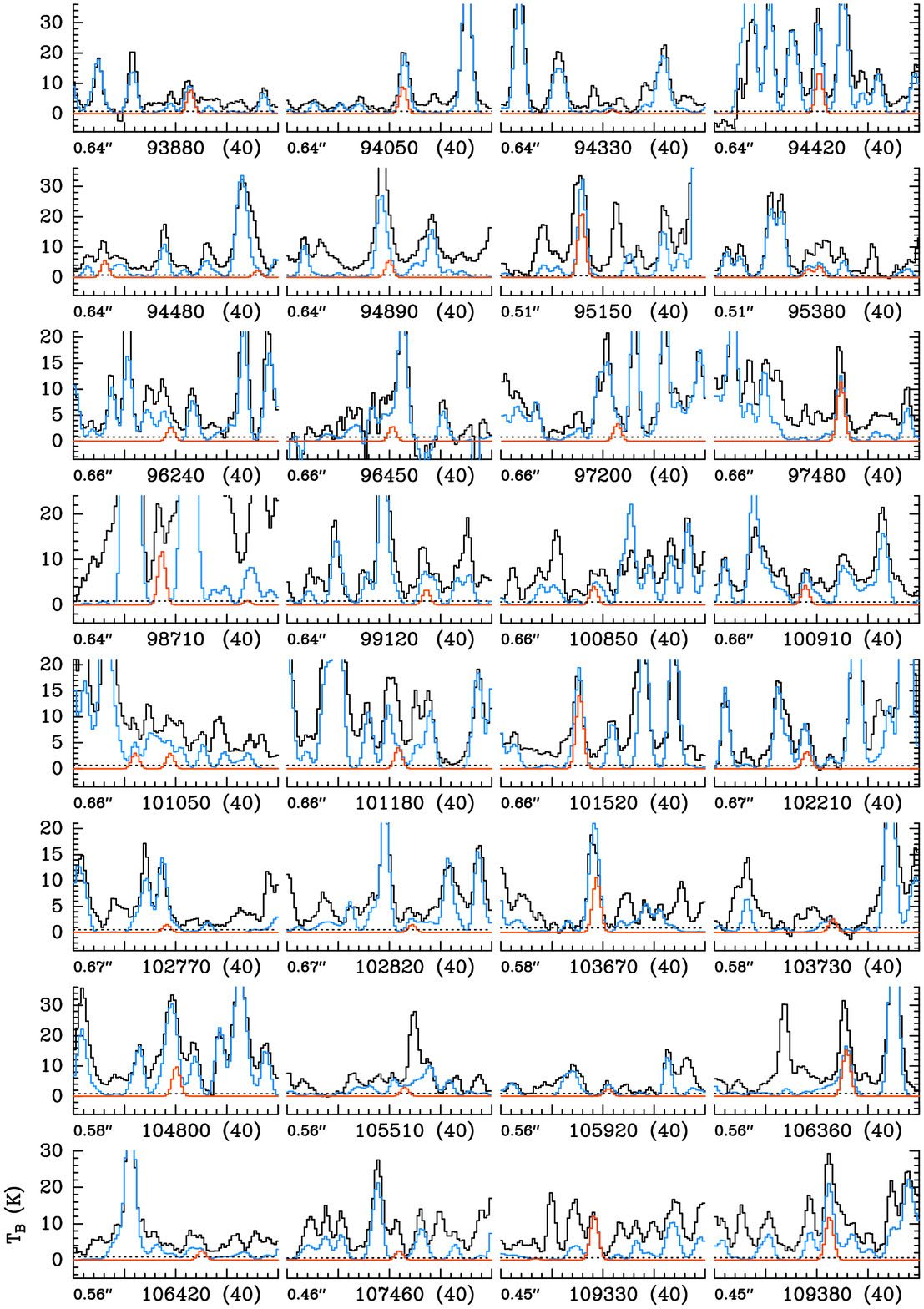}}}
\caption{continued.}
\end{figure*}

\begin{figure*}
\addtocounter{figure}{-1}
\centerline{\resizebox{0.9\hsize}{!}{\includegraphics[angle=0]{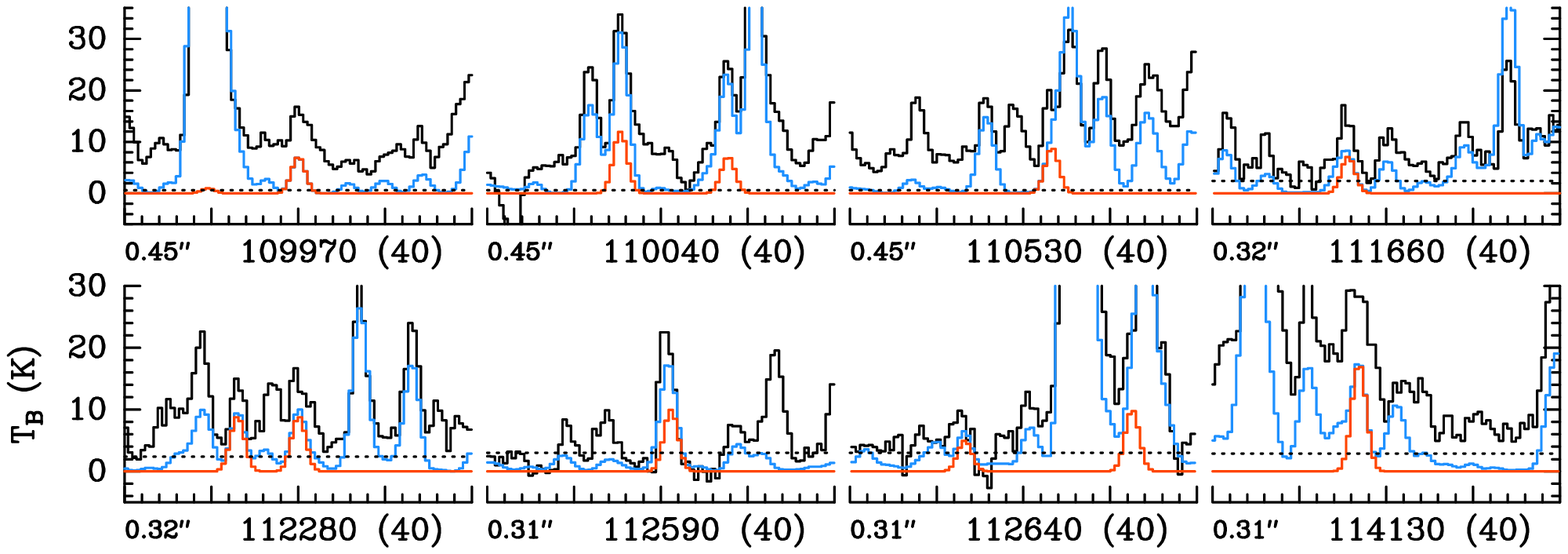}}}
\caption{continued.}
\end{figure*}

\begin{figure*}[!h]
\vspace{1.0cm}
\centerline{\resizebox{0.95\hsize}{!}{\includegraphics[angle=0]{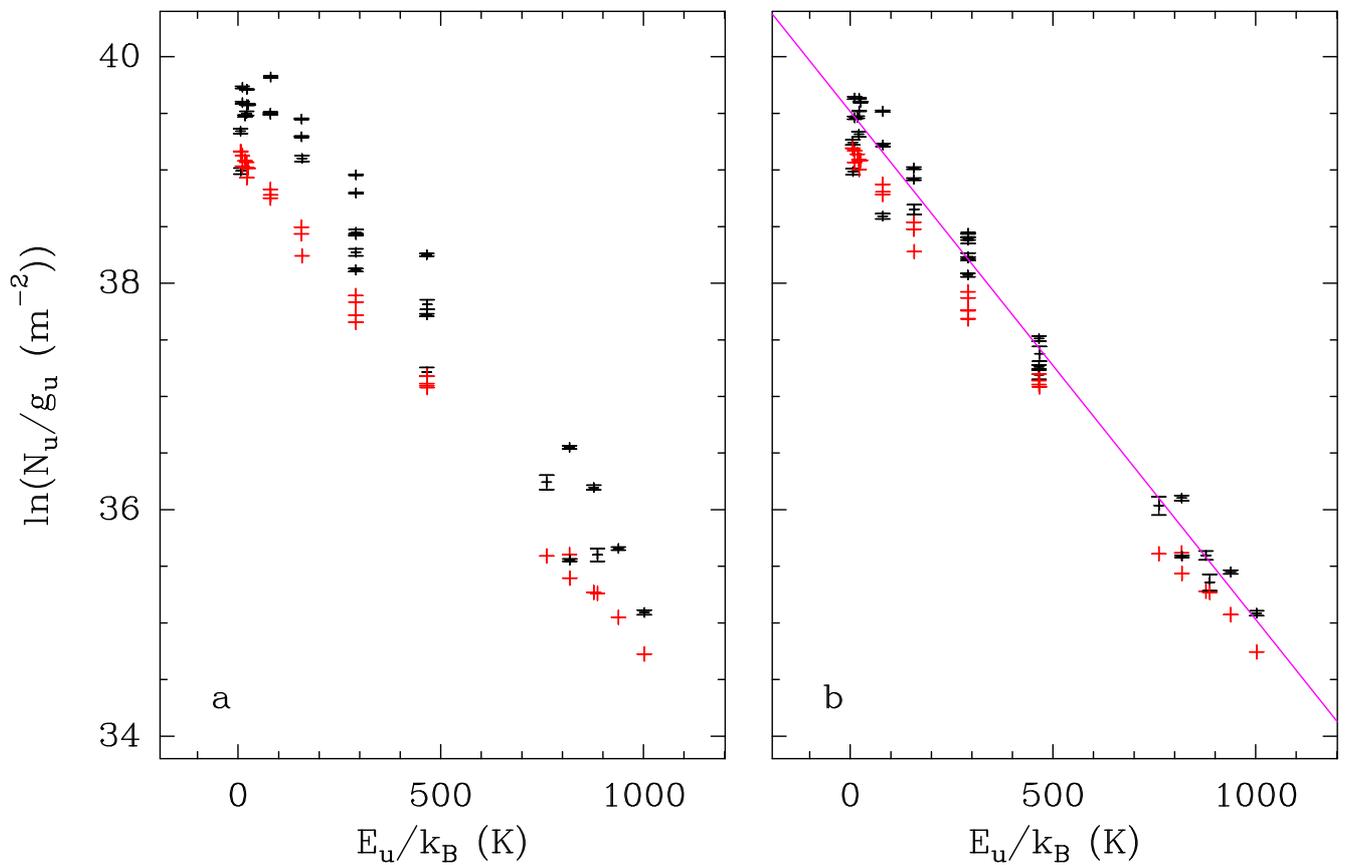}}}
\caption{Population diagram of NH$_2$CH$_3$ toward Sgr~B2(N1S). The
observed datapoints are shown in black while the synthetic
populations are shown in red. No correction is applied in panel {\bf a}.
In panel {\bf b}, the optical depth correction has been applied to both the
observed and synthetic populations and the contamination by all other
species included in the full model has been removed from the observed
datapoints. The purple line is a linear fit to the observed populations (in
linear-logarithmic space).
}
\label{f:popdiag_nh2ch3}
\end{figure*}

\end{appendix}

\end{document}